\documentclass[lengthcheck,onecolumn,pre,superscriptaddress,showpacs,floatfix,aps]{revtex4-2}

\usepackage[percent]{overpic}
\usepackage{fix-cm}
\usepackage{url}
\usepackage{subcaption}
\usepackage{amsmath,amssymb,amsfonts,latexsym,bm}
\usepackage{graphicx,epsfig}
\usepackage{fancyvrb}
\usepackage{booktabs}
\usepackage{soul}
\usepackage{algorithm}
\usepackage{algpseudocode}
\usepackage{ragged2e}
\usepackage[svgnames]{xcolor}
\usepackage[colorlinks,linkcolor=MediumBlue,citecolor=MediumBlue,urlcolor = MediumBlue]{hyperref}

\newcommand{\ve}[1]{\boldsymbol{#1}}
\newcommand{\ma}[1]{\ensuremath{\mathbb{#1}}}

\allowdisplaybreaks

\begin{document}

\title{Adaptive shape control for microswimmer navigation in turbulence}

\author{Jingran Qiu}
\affiliation{Department of Physics, Gothenburg University, 41296 Gothenburg, Sweden}
\author{Lorenzo Piro}\email{lorenzo.piro@roma2.infn.it}
\affiliation{Department of Physics \& INFN, Tor Vergata University of Rome, Via della Ricerca Scientifica 1, 00133 Rome, Italy}
\author{Luca Biferale}
\affiliation{Department of Physics \& INFN, Tor Vergata University of Rome, Via della Ricerca Scientifica 1, 00133 Rome, Italy}
\author{Massimo Cencini}
\affiliation{Istituto dei Sistemi Complessi, CNR, Via dei Taurini 19, 00185 Rome, Italy}
\affiliation{INFN ``Tor Vergata", Via della Ricerca Scientifica 1, 00133 Rome, Italy}
\author{Bernhard Mehlig}
\affiliation{Department of Physics, Gothenburg University, 41296 Gothenburg, Sweden}
\author{Kristian Gustavsson}
\affiliation{Department of Physics, Gothenburg University, 41296 Gothenburg, Sweden}

\date{\today}

\begin{abstract}
Navigation in turbulent environments is a fundamental challenge for biological and artificial microswimmers. While most existing studies focus on adapting motility or steering, the role of active morphological changes in navigation remains poorly explored. Here, we investigate a shape-changing spheroidal microswimmer tasked with maximising its displacement from an initial position in two-dimensional stochastic and turbulent flows. Using reinforcement learning (RL), the microswimmer learns to adapt its aspect ratio based on its orientation and local velocity-gradient signals.
The learned strategies outperform fixed-shape and short-time-optimal baselines across different flow regimes and remain effective when transferred from stochastic flows to fully resolved turbulence. Guided by the learned policies, we propose a minimal analytical model that captures the essential navigation mechanisms and reproduces the performance across flow regimes. These results show that adaptive morphology provides a robust and physically interpretable control paradigm for microswimmer navigation in complex flows.
\end{abstract}

\maketitle

\section{Introduction}

Navigation and exploration of microswimmers in turbulent and complex flow environments are fundamental tasks in both natural and artificial systems.
In nature, for example, phytoplankton and zooplankton live in a turbulent environment, spending significant effort in vertical migration and nutrient foraging~\citep{hays2003Reviewa,guasto2012Fluida}. 
Artificial microswimmers, by contrast, have shown promise for biomedical applications such as targeted drug delivery and microsurgery~\citep{huang2019Adaptive,Tsang2020}.

An intriguing scenario arises when microswimmers perceive and adapt to environmental cues. This capacity is common among many plankton species, which can sense and respond to environmental cues such as light (phototaxis)~\citep{jekely2008Mechanism}, chemical gradients (chemotaxis)~\citep{stocker2012Ecology}, and hydromechanical signals~\citep{kiorboe1999Predator}.
Beyond modulating motility, some plankton also alter their morphology in response to environmental changes. For instance, \citet{Sengupta_Carrara_Stocker_2017} showed that certain algae undergo shape transformations under turbulent conditions, thereby reversing their vertical migration by altering their stable orientation. Similarly, some diatoms form long chains to adjust their effective aspect ratio, thereby modifying their dynamics in turbulence and improving vertical migration efficiency~\citep{smayda2010Adaptations,borgnino2018,lovecchio2019Chain}.

In synthetic systems, artificial microswimmers must adapt to heterogeneous and time-dependent flow environments to achieve autonomous navigation. Despite substantial progress, this objective has not yet been realised. Advances in biohybrid microrobots have demonstrated adaptive and partially autonomous behaviour~\citep{ricotti2017Biohybrid,huang2019Adaptive,Tsang2020}. For example, \citet{huang2016Soft} developed a microswimmer with deformable flagella that enable controllable swimming gaits. However, fully autonomous microswimmers that operate without external control remain a major challenge, since adaptive capabilities are typically hard-coded during fabrication, which limits flexibility and practical applications.

The study of adaptive navigation strategies is therefore important not only for gaining insight into plankton ecology, but also for guiding the design of artificial microswimmers. In both settings, microswimmers must account for the local flow field to navigate efficiently. 
In recent years, the motion of microswimmers in turbulent environments has been investigated extensively. Modelling them as Lagrangian spheroidal particles swimming in their instantaneous direction explains phenomena such as small-scale clustering~\citep{durham2013Turbulence,gustavsson2016Preferential} and preferential alignment with flow structures~\citep{zhan2014Accumulation,borgnino2019Alignment}. Particle shape plays a central role in this context: it alters the hydrodynamic torque on the swimmer, modifying preferential orientation~\citep{borgnino2019Alignment,ma2022Reaching} and location in the flow~\citep{torney2007Transport,gustavsson2016Preferential,borgnino2018}. 
Because the rotational dynamics directly influences the swimming direction and stability, adaptive morphological changes may provide significant advantages for navigation in complex environments.

In this context, reinforcement learning has recently emerged as a powerful tool for discovering optimal strategies in complex environments~\citep{sutton2018Reinforcementa,mehlig2021machine}. Since the pioneering work of \citet{colabrese2017Flow}, reinforcement learning has been widely applied to the study of microswimmer navigation strategies~\citep{gustavsson2017Finding,alageshan2020Machine,gunnarson2021Learning,borra2022,xu2022Migration,zou2022Gait,qiu2022Navigation,qiu2022Active,mousavi2024Efficient,mousavi2025Short}. However, except for~\cite{colabreseSmartInertialParticles2018}, which considered density-changing microswimmers, most existing studies have focused on adaptive motility in response to environmental cues, such as swimming speed, steering direction, or locomotory gaits. In contrast, navigation strategies based on active deformation or morphological adaptation remain largely unexplored.

In this study, we use reinforcement learning to investigate the navigation strategy of shape-changing spheroidal microswimmers in two-dimensional stochastic and turbulent flows.
We consider the task in which a microswimmer needs to maximise its distance from its initial position over a fixed time horizon [Fig.~\ref{fig:swimmer}(a)]. This serves as a minimal model for ecologically relevant behaviours such as evasion from harmful regions~\citep{incze2001Changes}, and efficient exploration of nutrition-rich environments~\citep{bartumeus2003Helical}.
The microswimmer is assumed to sense its instantaneous orientation and local flow-velocity gradients, and to adapt its aspect ratio accordingly. 

Our main findings can be summarised as follows: \textbf{(i) Adaptive shape control enhances navigation.} The learned strategy consistently outperforms heuristic baselines, with the largest gains in slowly evolving flows. \textbf{(ii) A minimal analytical model captures the learned behaviour.} A reduced interpretable strategy distilled from the reinforcement-learning policy reproduces performance across flow regimes. \textbf{(iii) The learned adaptation mechanisms are robust and transferable.} Policies trained in stochastic flows transfer successfully to turbulent flows and retain high performance under moderate variations of the swimming parameters, indicating that the extracted mechanisms are universal rather than artefacts of a specific flow model.

Taken together, these results show that active shape modulation provides a physically interpretable and efficient control paradigm for microswimmer navigation in complex flows, revealing how morphological adaptation can be leveraged to exploit environmental structure over multiple timescales.

The paper is organised as follows. Section~\ref{sec:methods} introduces the microswimmer model, stochastic flow model, and reinforcement learning framework. 
Section~\ref{sec:baselines} defines the benchmark strategies.
Section~\ref{sec:results} analyses the strategies learned in different flow regimes, compares their performances with the benchmarks, explains the underlying physical mechanisms, and examines their transferability to direct numerical simulations (DNS) of turbulence. Finally, Section~\ref{sec:conclusions} summarises the main findings and outlines future directions.

\section{Methods}
\label{sec:methods}
\subsection{Microswimmer model}

\begin{figure}
\centering
\begin{overpic}[width=0.9\textwidth,percent]{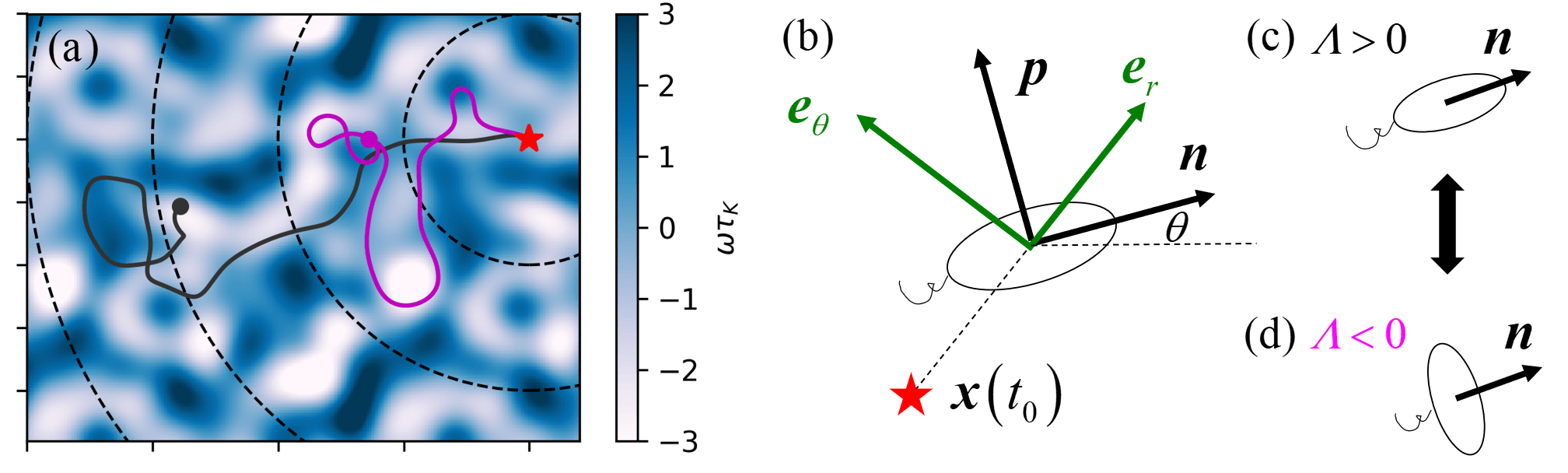}
\end{overpic}
    \caption{
    (a) Microswimmer navigation maximizing distance from the initial position $\ve x(t_0)$ (red star).
    Equidistant black dashed circles denote radial distances.
    Two example trajectories of microswimmers following Eq.~(\ref{eq:eom}) with constant shape factor, evolving in a frozen realization of the stochastic model (magenta: $\Lambda=-0.98$; black: $\Lambda=0.98$).
    (b) Schematic of a spheroidal microswimmer and its reference frame. (c,d) Microswimmers can adjust their aspect ratio, becoming (c) prolate, $\Lambda>0$, or (d) oblate, $\Lambda<0$.
    }
    \label{fig:swimmer}
\end{figure}

We model microswimmers as self-propelled spheroidal particles advected by a fluid flow [Fig. 1(b)]. Each swimmer is characterised by its position $\ve x(t)$ and by a unit orientation vector $\ve n(t)$ aligned with its symmetry axis. The shape is parametrised by the factor $\Lambda= (\lambda^2-1)/(\lambda^2+1)$, where $\lambda$ is the spheroid aspect ratio. We assume that microswimmers can actively deform and control $\Lambda$, as illustrated in Fig.~\ref{fig:swimmer}(c, d).

Assuming small microswimmers and negligible inertia, their dynamics is governed by the standard model for elongated particles in viscous flows~\citep{kessler1987,durham2009Disruption,qiu2022Navigation,mousavi2024Efficient}:
\begin{subequations}
\begin{align}
    \dot{\ve x} = & \ve u + v_{\rm s} \ve n\,,     \label{eq:tran}                                                \\
    \dot{\ve n} = & \frac{1}{2}\ve{\omega}\times \ve n + \Lambda \left( \ma{I} - \ve n \ve n\right)\cdot \ma{S} \cdot \ve n\,.
    \label{eq:rot}
\end{align}
\label{eq:eom}
\end{subequations}
Here $\ve u(\ve x,t)$ denotes the fluid velocity evaluated at the microswimmer position, $v_{\rm s}$ denotes swimming speed, $\ve \omega = \nabla \times \ve u$ is the vorticity, $\ma S = \tfrac{1}{2}(\nabla \ve u + \nabla \ve u^{T})$ is the strain-rate tensor, and $\ma I$ is the identity matrix.

The translational dynamics in Eq.~(\ref{eq:tran}) consists of passive advection by the ambient flow and active propulsion along the swimmer orientation $\ve n$. 
The rotational dynamics in Eq.~(\ref{eq:rot}) describes the rotation of a spheroid by velocity gradients in viscous flow~\citep{jeffery1922}. The contribution of the strain-rate tensor $\ma S$ is proportional to the shape factor $\Lambda$. Consequently, particle shape strongly impacts microswimmer dispersion and transport~\citep{khurana2012Interactions,gustavsson2016Preferential,piro2024Energetica}. 
In particular, swimmers with different $\Lambda$ can exhibit qualitatively distinct trajectories within the same flow field [Fig.~\ref{fig:swimmer}(a)], indicating that active modulation of $\Lambda$ may serve as an effective navigation mechanism.

We focus on a two-dimensional flow in which the microswimmer orientation is confined to the flow plane. In this case, the orientation dynamics in Eq.~(\ref{eq:rot}) simplifies to 
\begin{equation}
    \dot{\theta}=\frac{1}{2}\omega + \Lambda \ve{p}\cdot\ma{S}\cdot \ve{n}\,,
    \label{eq:thetaDot}
\end{equation}
where $\theta$ is the angle of $\ve n$, $\omega$ is the out-of-plane vorticity, and $\ve p$ is the in-plane unit vector perpendicular to $\ve n$ [Fig.~\ref{fig:swimmer}(b)].

\subsection{Stochastic flow model}
\label{sec:flowmodel}
To model the turbulent fluctuations, we use a two-dimensional, time-dependent, isotropic stochastic flow model~\citep{bec2024Statistical}. The velocity field of the flow is determined by $\ve u(\ve x,t) = (\partial \psi/\partial y, - \partial \psi/\partial x)$. Following \citet{mousavi2024Efficient}, the stream function $\psi(\ve x,t)$ is the superposition of $M$ time-independent components $\psi_i(\ve x)$, with time-dependent coefficients $c_i(t)$.
The components follow the spatial correlation,
\begin{equation}
    \langle\psi_i(\ve x)\psi_j(\ve x^\prime)\rangle = \frac{1}{2}\delta_{ij}u_{\rm rms}^2 \ell_{\rm f}^2\exp{\left[-\frac{(\ve x-\ve x^\prime)^2}{2\ell_{\rm f}^2}\right]}\,,
    \label{eq:flowmodel}
\end{equation}
where $u_{\rm rms} = \sqrt{\langle \ve u \cdot \ve u \rangle}$ is the root-mean-square flow velocity, and $\ell_{\rm f}$ is the flow correlation length scale. 
The time-dependent coefficients $c_i(t)$ follow an Ornstein-Uhlenbeck process with zero mean and time correlation
\begin{equation}
    \langle c_i(t)c_j(t^\prime)\rangle  = \frac{\delta_{ij}}{M} \exp\left({-\frac{|t-t^\prime|}{\tau_{\rm f}}}\right)\,,
\end{equation}
where $\tau_{\rm f}$ is the flow correlation time.

The stochastic flow is characterised by a single dimensionless parameter, the Kubo number ${\rm Ku}=\tau_{\rm f} u_{\rm rms}/ \ell_{\rm f}$, which compares the flow correlation time $\tau_{\rm f}$ to the advective timescale $\ell_{\rm f}/u_{\rm rms}$. Large ${\rm Ku}$ corresponds to slowly evolving flows, approaching a time-independent limit as ${\rm Ku} \to \infty$, while ${\rm Ku}\to0$ yields temporally uncorrelated flows with rapid fluctuations.
This stochastic flow model reproduces essential statistical properties of small-scale velocity-gradient fluctuations in turbulent flows, showing good agreement with DNS for both inertial particles~\citep{bec2024Statistical} and microswimmers~\citep{gustavsson2016Preferential,borgnino2019Alignment,qiu2022Active,mousavi2024Efficient,mousavi2025Short}, especially in the regime of ${\rm Ku}\sim 10$.

In this work, we vary the Kubo number by adjusting the flow correlation time $\tau_{\rm f}$, while keeping the root-mean-square flow velocity $u_{\rm rms}$ and length scales $\ell_f$ fixed. To match the stochastic flow to turbulence, we take $u_{\rm rms}$ as the characteristic velocity scale and the Kolmogorov timescale, $\tau_{\rm K} = \langle \tfrac{\partial u_i}{\partial x_j}\tfrac{\partial u_i}{\partial x_j}  \rangle^{-1/2}$, as the characteristic flow time scale.

\subsection{Reinforcement learning}
\label{sec:methods:rl}

The goal is to maximise the swimmer's displacement from its initial position over a prescribed time horizon.
To identify an effective control strategy, we employ Proximal Policy Optimisation (PPO)~\citep{schulman2017Proximal}, a deep reinforcement-learning algorithm designed for continuous state and action spaces. The policy network maps the swimmer state to a continuous control value corresponding to the shape factor $\Lambda$. 
We assume that the microswimmer senses its state at discrete times separated by an update interval $T_{\rm u}$, and updates its shape $\Lambda$ accordingly.
The state is composed of local signals: swimmer orientation, strain-rate $\ma S$, and vorticity $\omega$.

Due to the radial symmetry about the initial position, we decompose the orientation $\ve n$ in the polar basis spanned by the radial unit vector $\ve e_r = \ve x/|\ve x|$ and the azimuthal unit vector $\ve e_\theta$ [Fig.~\ref{fig:swimmer}(b)]. 
The projected components $n_r = \ve n \cdot \ve e_r$ and $n_\theta = \ve n \cdot \ve e_\theta$ are included in the state. The former indicates whether the swimmer is heading away from or toward the origin, while the latter provides information on which direction to rotate. 

Consistent with previous studies~\citep{kiorboe1999Predator,qiu2022Navigation,qiu2022Active,borra2022}, we assume that velocity gradients are measured in the body frame ${\ve n,\ve p}$ [Fig.~\ref{fig:swimmer}(b)]. In two-dimensional incompressible flow, the strain-rate tensor has two independent components, $S_{nn} = \ve n \cdot \ma S \cdot \ve n$ and $S_{np} = \ve n \cdot \ma S \cdot \ve p$, while the vorticity reduces to the scalar $\omega$. 
These quantities are included in the state because they directly govern the rotational dynamics~(\ref{eq:rot}), encode the local flow topology, and have been shown to be central to efficient microswimmer navigation in diverse  settings~\citep{jeffery1922,gunnarson2021Learning,monthiller2022Surfing,qiu2022Active,mousavi2024Efficient,mousavi2025Short}.

To optimize swimmer displacement from the initial location $\ve x(t_0)$, we introduce the step-wise reward
\begin{equation}
    r_i = |\ve x(t_i) - \ve x(t_0)| - |\ve x(t_{i-1}) - \ve x(t_0)|\,,
    \label{eq:reward}
\end{equation}
where $\ve x(t_i)$ denotes the microswimmer position at the discrete update time $t_i = i T_{\rm u}$. Since rewards are evaluated at fixed time intervals, maximising the cumulative reward is equivalent to maximising the radial velocity away from $\ve x(t_0)$.
Noting that $\sum_{i=1}^N r_i = |\ve x(t_N) - \ve x(t_0)|$, we define the performance index as 
\begin{equation}
    \label{eq:reward2}
    \hat{R} = |\ve x(t_N) - \ve x(t_0)|/R^{(0)}_{\rm max} \, ,
\end{equation}
where $R^{(0)}_{\rm max}\equiv v_{\rm s}T$ is the distance travelled over the duration of an episode $T = N T_{\rm u}$ by a swimmer moving in a straight line from $\ve x(t_0)$ at speed $v_{\rm s}$ in quiescent flow.

Training is performed over a large number of episodes. At the start of each episode, the flow coefficients $c_i$ and the swimmer's initial position and orientation are randomly sampled. The swimming dynamics is then numerically integrated over the episode duration $T$ (see Appendix~\ref{app:simulations}) following a policy fixed for one episode. The trajectories collected during each episode are used to update the policy parameters using the PPO algorithm (see Appendix~\ref{app:RL}). 
Convergence is assumed when the performance index $\hat{R}$ plateaus as the number of episode increases.

\begin{table}
   \begin{center}
    \begin{tabular}{lll}
        Description                      & Parameter           & Value                                                   \\ \hline
        Swimming speed                   & $v_{\rm s}$            &  $0.707u_{\rm rms}$                                             \\
        Shape factor & $\Lambda$ & $-0.98\le \Lambda \le 0.98$ \\
                Kubo number    & ${\rm Ku}$                 & 0.1, 1, 10, $10^7$\\     
        Signals                          &                     & $n_r, n_\theta, S_{nn}, S_{np}, \omega $ \\

        Update interval   & $T_{\rm u}$ & 0.1$\tau_{\rm K}$                                            \\
        Episode duration       & $T$                 & $200 \tau_{\rm K}$                                            \\               
    \end{tabular}
     \caption{Summary of the simulation parameters and of the states (signals) available to the swimmer. 
     }
    \label{tab:para}
    \end{center}
\end{table}

The parameters of numerical simulations and training are summarised in Table~\ref{tab:para}.
The swimming speed is set to $v_{\rm s}=0.707u_{\rm rms}$, corresponding to microswimmers with a moderate propulsion speed compared to the fluid flow.
The shape factor $\Lambda$ is constrained between $-0.98$ and $0.98$, corresponding to aspect ratios between $0.1$ and $10$ [Fig.~\ref{fig:swimmer}(c,d)].
We consider four Kubo numbers, ${\rm Ku} = 0.1$, $1$, $10$, and $10^7$, covering regimes from rapidly decorrelating to effectively steady flows. 
The episode duration is fixed to $T=200\tau_{\rm K}$, ensuring that initial transients are negligible. 
The control update interval is set to $T_{\rm u}=0.1\tau_{\rm K}$, allowing the microswimmer to respond to temporal variations in the local flow.

For each value of the Kubo number ${\rm Ku}$, the training procedure is repeated three times using different random seeds. In all cases, the performance of the learned smart strategy increases with the number of episodes and saturates after approximately $5000$ episodes, indicating convergence (Fig.~\ref{fig:trainingCurves}). The independent runs yield quantitatively similar performance; consequently, we report results from a single representative training for each ${\rm Ku}$. The robustness of the smart strategies to stochastic perturbations to the microswimmer dynamics and to variations in training parameters (such as the update interval $T_{\rm u}$, the swimming speed $v_{\rm s}$, and the episode duration $T$) is documented in Appendix~\ref{app:robustness}.

\section{Baseline strategies}
\label{sec:baselines}

To assess the quality of the learned policies, we compare their performances with two baseline strategies: one based on optimal control and one based on short-time optimization.

\subsection{Naive strategy}
An optimal control formulation of the navigation problem is obtained by the cost functional
\begin{equation}
    \mathcal{C} = - |\bm{x}(t)-\bm{x}(t_0)|^2 \,,
    \label{eq:cost}
\end{equation}
so that minimizing $\mathcal{C}$ is equivalent to maximizing the distance from the initial position in the shortest time. This choice is consistent with the problem definition and with the reward in Eq.~\eqref{eq:reward}. 
Applying Pontryagin’s minimum principle~\citep{bryson} to the associated Hamiltonian shows that the optimal strategy is to maintain a constant shape factor $\Lambda = -1$ (see Appendix~\ref{app:OCT}).
The control problem can, in fact, be mapped to Zermelo’s minimum-time navigation problem~\citep{zermelo} with an infinitely distant target. 
In this framework, the angular dynamics for $\Lambda = -1$ in Eq.~(\ref{eq:rot}) coincides with that obtained from Zermelo’s formulation~\citep{piro2024Energetica}. 
Consequently, provided the initial orientation is appropriately chosen, swimmers with $\Lambda=-1$ reach any admissible target point, including the point farthest from the origin, faster than microswimmers with other values of $\Lambda$.
However, in turbulent flows, trajectories with $\Lambda=-1$ are extremely sensitive to initial conditions and most become trapped in high-vorticity regions~\citep{biferale2019chaos,piro2022frontiers}.
Identifying the initial orientation that yields optimal long-time performance is therefore impractical. 
Evaluating trajectories for a large number of initial conditions with $\Lambda \in [-0.98, 0.98]$, we find instead that elongated microswimmers ($\Lambda = 0.98$) achieve larger distances on average and yield the highest mean performance index $\hat{R}$. For this reason, we adopt $\Lambda = 0.98$ microswimmers as a naive baseline in the following analysis.

\subsection{Short-time optimization (STO) strategy}
A more refined baseline is based on the idea of short-time optimization, which has demonstrated its capacity to solve the optimal steering strategy of a microswimmer performing vertical migration~\citep{monthiller2022Surfing} or escaping high-strain regions~\citep{mousavi2025Short}.
In our setup, we derive the STO by maximizing the reward in Eq.~(\ref{eq:reward}) over a short time interval $\delta t$.
Consider a microswimmer at position $\ve x_t$ at time $t$ and choose $\Lambda$ to maximise the incremental reward until $t+\delta t$, i.e. maximise $|\ve x(t+\delta t)-\ve x(0)|$. Shifting the coordinate system so that $\ve x(0)=0$, we have
\begin{align}
|\ve x(t+\delta t)|=
|\ve x|+\delta t\,\dot{\ve x}\cdot\ve e_r+\frac{1}{2}\delta t^2\Big[\ddot{\ve x}\cdot\ve e_r+\frac{1}{|\ve x|}(\dot{\ve x}^2-\big(\dot{\ve x}\cdot\ve e_r)^2\big)\Big]+O(\delta t^3)\,,
\label{eq:reward_increase_dt}
\end{align}
where the right-hand side is evaluated at time $t$.
Using the dynamics in Eq.~(\ref{eq:eom}), the only shape-dependent contribution to Eq.~(\ref{eq:reward_increase_dt}) is contained in $\frac{1}{2}\delta t^2\ddot{\ve x}\cdot\ve e_r$:
\begin{align}
|\ve x(t+\delta t)|=
\frac{v_{\rm s}\Lambda\delta t^2}{2}\ve e_r\cdot(\ma I-\ve n\ve n)\cdot\ma S(\ve x(t),t)\cdot\ve n
+\mbox{$\Lambda$-independent terms}+O(\delta t^3)\,.
\end{align}
Here $\ve e_r\cdot(\ma I-\ve n\ve n)$ is the projection of $\ve e_r$ on the subspace perpendicular to $\ve n$.
In two spatial dimensions, this projection becomes $(\ve e_r\cdot\ve p)\ve p=-n_\theta\ve p$, implying that $|\ve x(t+\delta t)|$ in this short-time expansion is maximised by choosing 
\begin{align}
\Lambda=-\Lambda_{\rm max}{\rm sign}(n_\theta(t))\,{\rm sign}(S_{np}(\ve x(t),t))\,.
\label{eq:sto}
\end{align}

In conclusion, short-time optimisation is achieved by taking $|\Lambda|$ constant over $\delta t$ and as large as allowed ($\Lambda_{\rm max}$), with the sign set by the local strain and microswimmer orientation.
If this strategy is followed and updated continuously, the angular dynamics~(\ref{eq:thetaDot}) becomes
\begin{align}
\dot\theta&=\frac{1}{2}\omega(\ve x(t),t)-\Lambda_{\rm max}{\rm sign}(n_\theta(t))|S_{np}(\ve x(t),t)|\,.
\label{eq:thetaDot_STO}
\end{align}
The mechanism can be understood as follows.
Neglecting $\omega(\ve x_t,t)$, the angular dynamics has fixed points at $n^*_\theta=0$, which corresponds to $\ve n^*=\pm \ve e_r(t)$.
The sign chosen in Eq.~(\ref{eq:sto}) is such that $\ve{n}^*=+\ve e_r(t)$ is stable. This deformation causes the strain to rotate the microswimmer away from its initial position, thereby maximizing its radial swimming velocity.
In simulations, we set $\Lambda_{\rm max}=0.98$ and update $\Lambda$ using the same update interval $T_{\rm u}$ as in the smart strategy, ensuring a fair comparison of performance.

\section{Results}
\label{sec:results}
We first compare smart navigation strategies learned via reinforcement learning with naive and short-time-optimisation (STO) baselines across flow regimes. We then analyse the smart strategies, identifying the physical mechanisms that govern their performances in rapidly fluctuating (${\rm Ku}\to0$) and steady (${\rm Ku}\to\infty$) flows, and how these mechanisms interpolate at intermediate Kubo numbers. From this, we distil a minimal analytical model that captures the essential ingredients of efficient navigation. Finally, we evaluate the smart strategy and the minimal model in 2D DNS of turbulence in the direct enstrophy-cascade regime.

\subsection{Performance comparison across flow regimes}

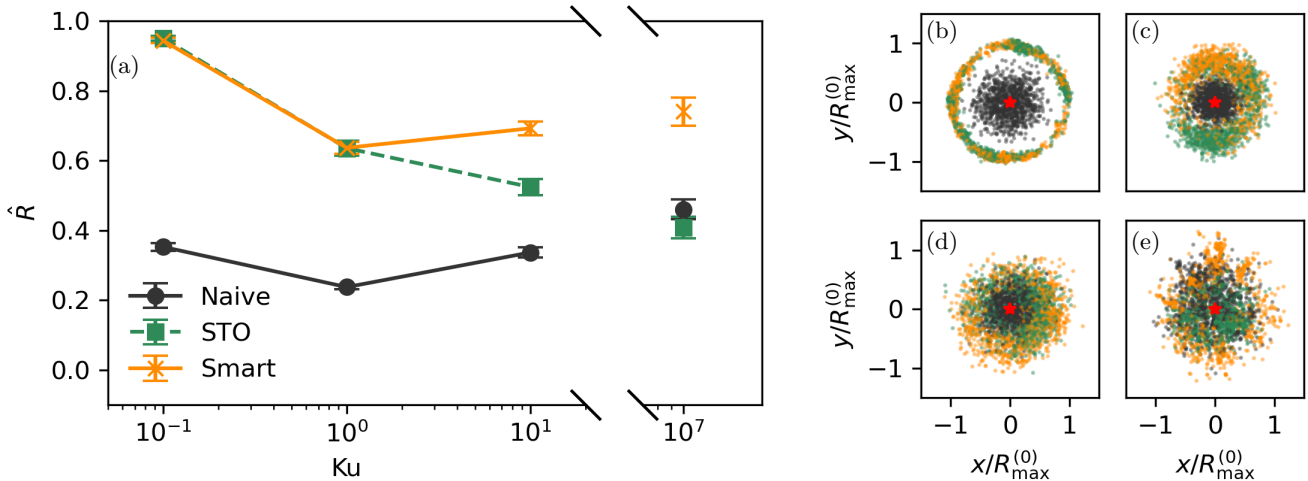
\begin{figure}
    \centering
    \begin{minipage}[t]{0.6\textwidth}
        \centering
        \begin{overpic}[width=\linewidth,percent,grid=False]{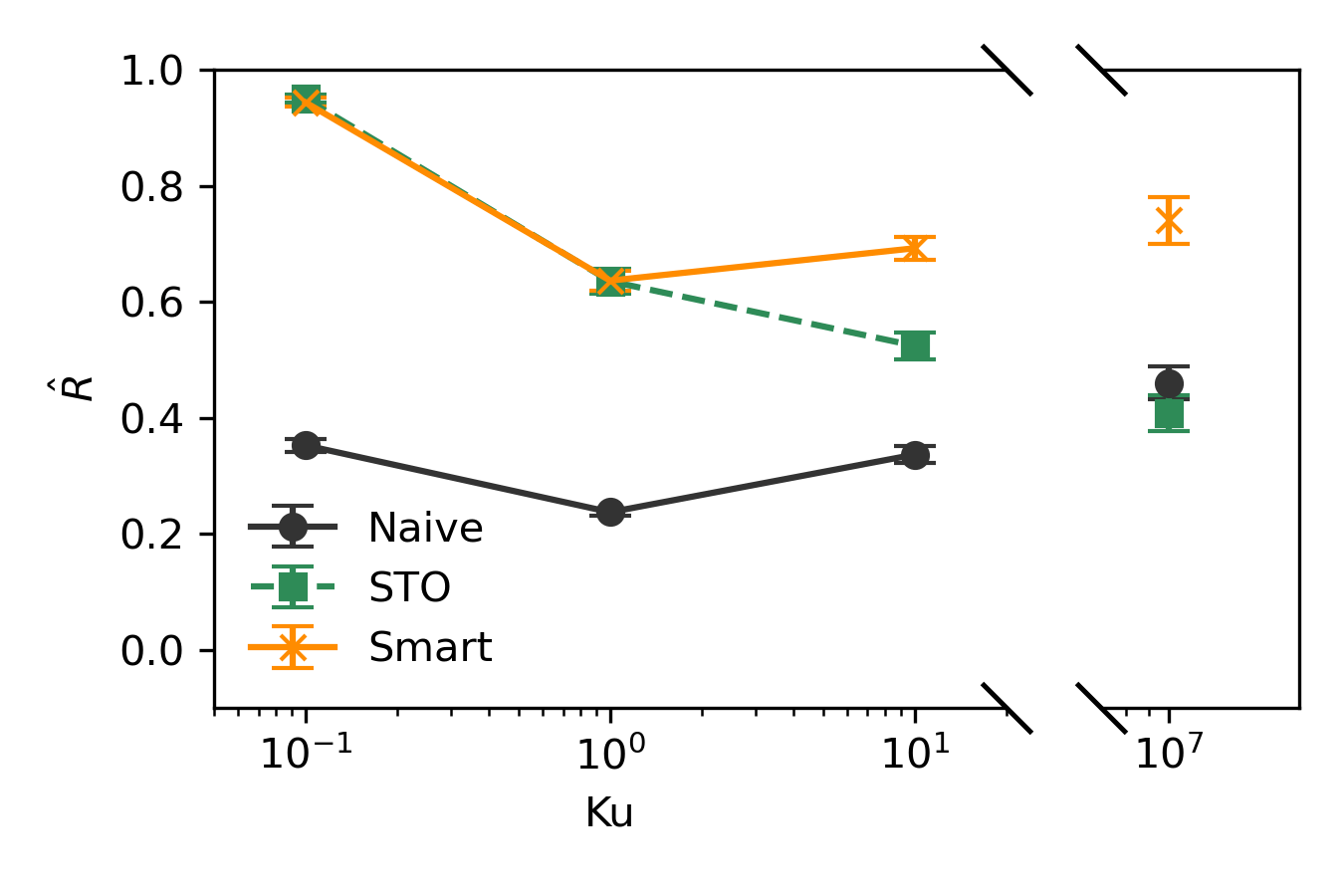}

            \put(16, 55){(a)} 
        \end{overpic}
    \end{minipage}%
    \hfill 
    \begin{minipage}[t]{0.4\textwidth}
        \centering
        \begin{overpic}[width=\linewidth,percent,grid=False]{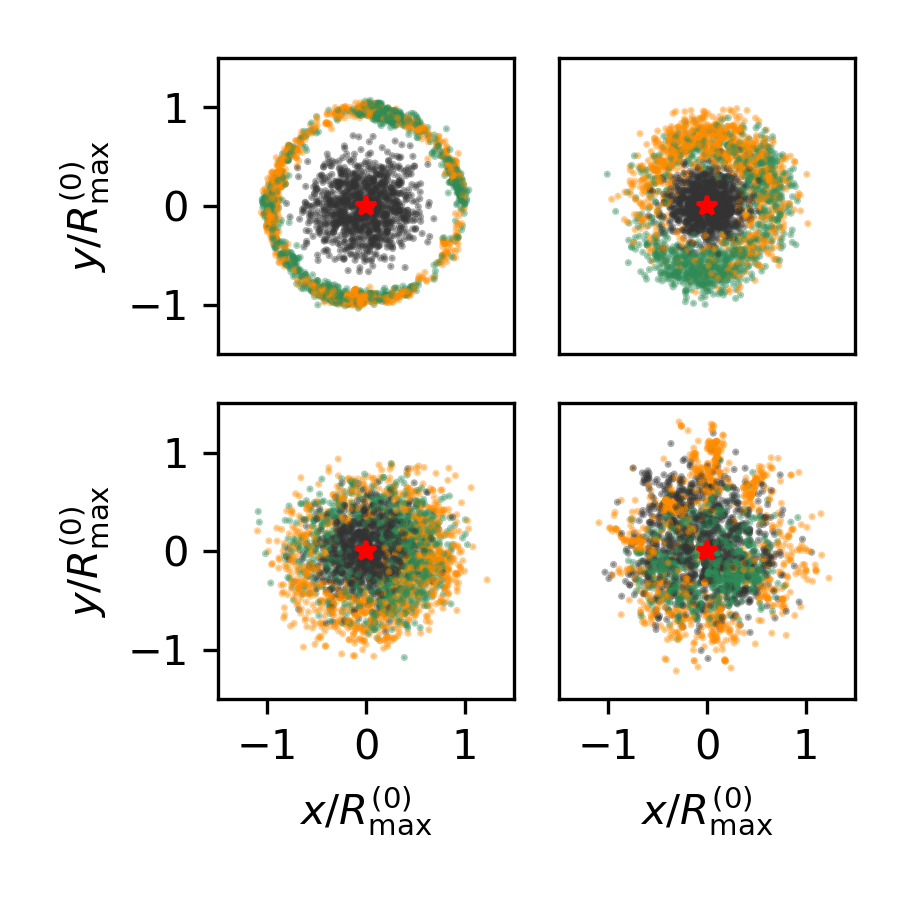}
            \put(25, 89){(b)} 
            \put(63, 89){(c)} 
            \put(25, 50){(d)} 
            \put(63, 50){(e)} 
        \end{overpic}
    \end{minipage}
    \caption{Comparison across flow regimes of reinforcement-learned smart strategy (orange), naive strategy [$\Lambda=0.98$] (black), and the STO strategy [Eq.~(\ref{eq:sto})] (green). (a) Normalised performance index $\hat{R}$ [Eq.~\eqref{eq:reward2}] against Kubo number ${\rm Ku}$. Error bars show the standard deviation of $\hat{R}$ over 100 flow realizations with $10^3$ trajectories each. (b) to (e): Final positions of $10^3$ microswimmers starting from the origin (red star) for (b) ${\rm Ku}=0.1$, (c) ${\rm Ku}=1$, (d) ${\rm Ku}=10$, (e) ${\rm Ku}=10^7$.}
    \label{fig:smartPerformance}
\end{figure}

Figure~\ref{fig:smartPerformance}(a) shows the maximal normalised performance index $\hat{R}$ versus Kubo number for different control strategies. The STO baseline performs best at small ${\rm Ku}$, with $\hat{R}$ decreasing monotonically as ${\rm Ku}$ increases. 
For small ${\rm Ku}$, flow gradients decorrelate rapidly, so reactive shape deformations suffice to maintain favourable orientations without long-term planning.
As flow persistence increases, however, this short-term strategy becomes progressively less effective.

The naive (constant shape) baseline shows a non-monotonic dependence on ${\rm Ku}$ and performs worse than the STO strategy except at very large ${\rm Ku}$ (${\rm Ku}=10^7$), where the flow is effectively steady, causing the STO strategy to fail as explained above. 
In contrast, the smart strategy obtained via RL based on all signals ($n_r$, $n_\theta$, $S_{nn}$, $S_{np}$, $\omega$) maintains high performance across all values of ${\rm Ku}$. For ${\rm Ku}\le1$, it matches the STO performance, indicating little room for improvement in rapidly fluctuating flows where STO is essentially optimal. At larger ${\rm Ku}$, however, it significantly outperforms both baselines: at ${\rm Ku}=10$, the displacement is approximately $30\%$ larger than STO, and in the quasi-steady limit (${\rm Ku}=10^7$) the gain exceeds $50\%$ relative to both STO and the naive strategies. These improvements demonstrate that the smart policy exploits long-term flow correlations that are inaccessible to strategies relying solely on instantaneous information.

Figure~\ref{fig:smartPerformance}(b–e) shows strategy performance via the final positions reached at different Kubo numbers.
The trends mirror those of $\hat R$: for large ${\rm Ku}$, smart microswimmers (orange) travel substantially farther than STO and naive swimmers, whereas for small ${\rm Ku}$, STO performs comparably to the smart strategy. Representative trajectories are provided in supplementary videos (Appendix~\ref{app:videos}).

Taken together, these results show that the reinforcement-learning strategy enables faster escape from the initial position, particularly in slowly evolving flows. The non-monotonic dependence of the normalised performance index on ${\rm Ku}$ indicates that distinct mechanisms underlie efficient navigation in rapidly fluctuating and steady flows. In the following sections, we analyse these mechanisms and their dependence on the flow regime.

\subsection{Smart strategy in rapidly fluctuating flow \texorpdfstring{$\rm Ku\to0$}{Ku→0}: STO recovery}
\label{sec:results:smallKu}
\begin{figure}
    \centering
    \begin{overpic}[width=1.0\textwidth,percent]{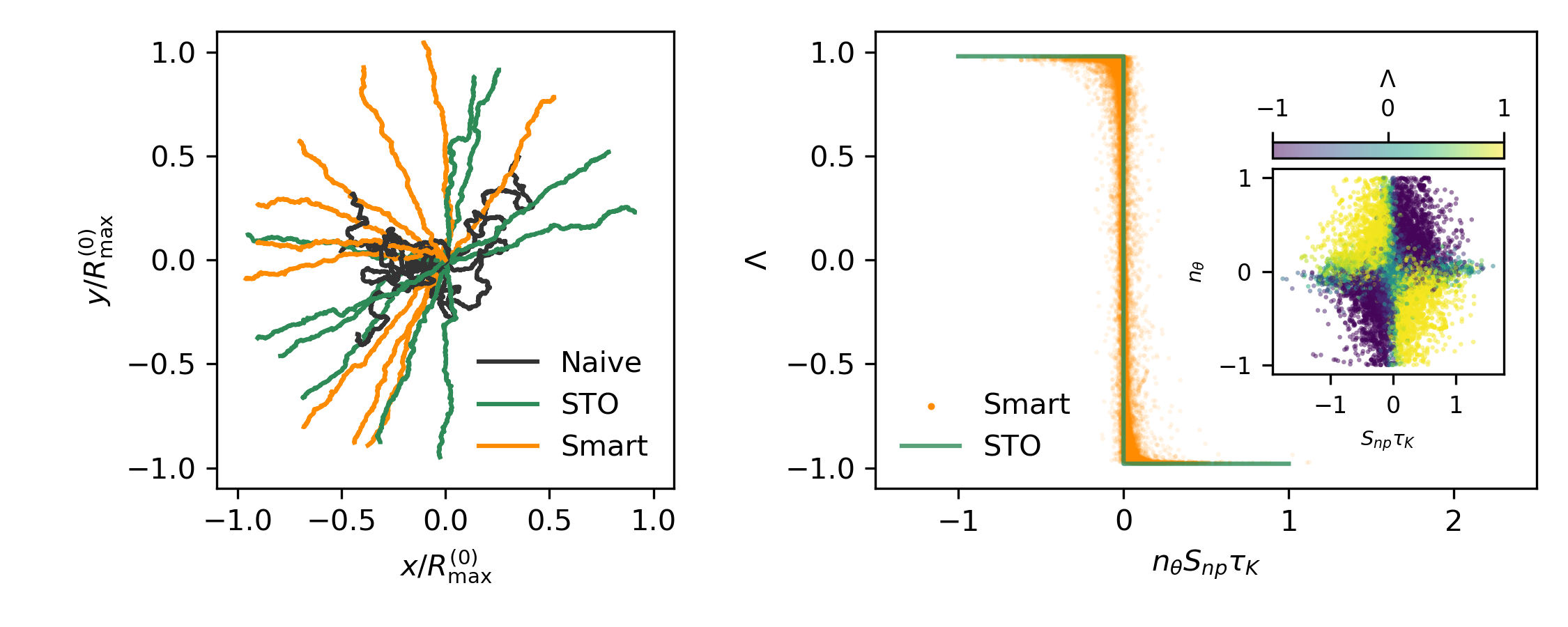}
        \put(15,35){(a)}
        \put(57,35){(b)}
    \end{overpic}
    \caption{(a) Trajectories of microswimmers using different strategies at ${\rm Ku}=0.1$: smart (orange), naive [$\Lambda=0.98$] (black), and STO [Eq.~(\ref{eq:sto})] (green). (b) Scatter plot of $\Lambda$ against $n_{\theta} S_{np}$ for the smart and STO strategies. Inset: $\Lambda$ against $n_\theta$ and $S_{np}$ for the smart strategy, based on $10^3$ trajectories.}
    \label{fig:ku0.1}
\end{figure}

In the rapidly fluctuating regime (${\rm Ku} = 0.1$), the smart strategy effectively coincides with STO.
Figure~\ref{fig:smartPerformance}(a) shows identical performance, and the trajectories are qualitatively the same: nearly straight paths directed away from the initial position [Fig.~\ref{fig:ku0.1}(a)]. 
Moreover, the learned control $\Lambda$ for all signals ($n_r$, $n_\theta$, $S_{nn}$, $S_{np}$, $\omega$) collapses onto the STO prediction [Eq.~(\ref{eq:sto})], when plotted against $n_\theta S_{np}$ [Fig.~\ref{fig:ku0.1}(b)].
Training using only $n_\theta$ and $S_{np}$ yields essentially the same behaviour as training with the full state vector, confirming that RL recovers STO as the optimal strategy at small ${\rm Ku}$ (data not shown).

The effectiveness of STO at small ${\rm Ku}$ follows from the rapid temporal decorrelation of the flow. As velocity gradients fluctuate quickly, long-term planning provides little advantage, and short-time optimisation approximates the globally optimal strategy. Consequently, the smart strategy does not outperform STO.

In the rapidly fluctuating regime, STO trajectories are nearly straight.
The effective STO dynamics in Eq.~(\ref{eq:thetaDot_STO}) breaks symmetry, making swimmers on average align with the radial direction for all values of ${\rm Ku}$. However, vorticity causes fluctuations around this direction, and for small ${\rm Ku}$ it acts as white noise on the otherwise stable orientation.
The orientation change over a time interval $\tau$ due to vorticity is
\begin{equation}
    \Delta \theta = \int_0^{\tau} \omega {\rm d}t\,.
\end{equation}
In the limit $\tau\gg\tau_{\rm f}$, this yields the diffusive scaling $(\Delta \theta)^2\approx 2\omega_{\rm rms}^2 \tau_{\rm f} \tau$. 
A shorter correlation time $\tau_{\rm f}$ therefore reduces orientation perturbations over a fixed time interval, helping the microswimmer remain close to the optimal direction. In this regime, short-time optimisation is sufficient: it captures the essential physics of efficient microswimmer navigation.

\subsection{Smart strategy in steady flow \texorpdfstring{$\rm Ku\to\infty$}{Ku→inf}: correlation among \texorpdfstring{$n_r$, $S_{np}$, and $\omega$}{nr,Snp,omega}}
\label{sec:results:largeKu}

\begin{figure}
    \centering
    \begin{overpic}[width=1\linewidth]{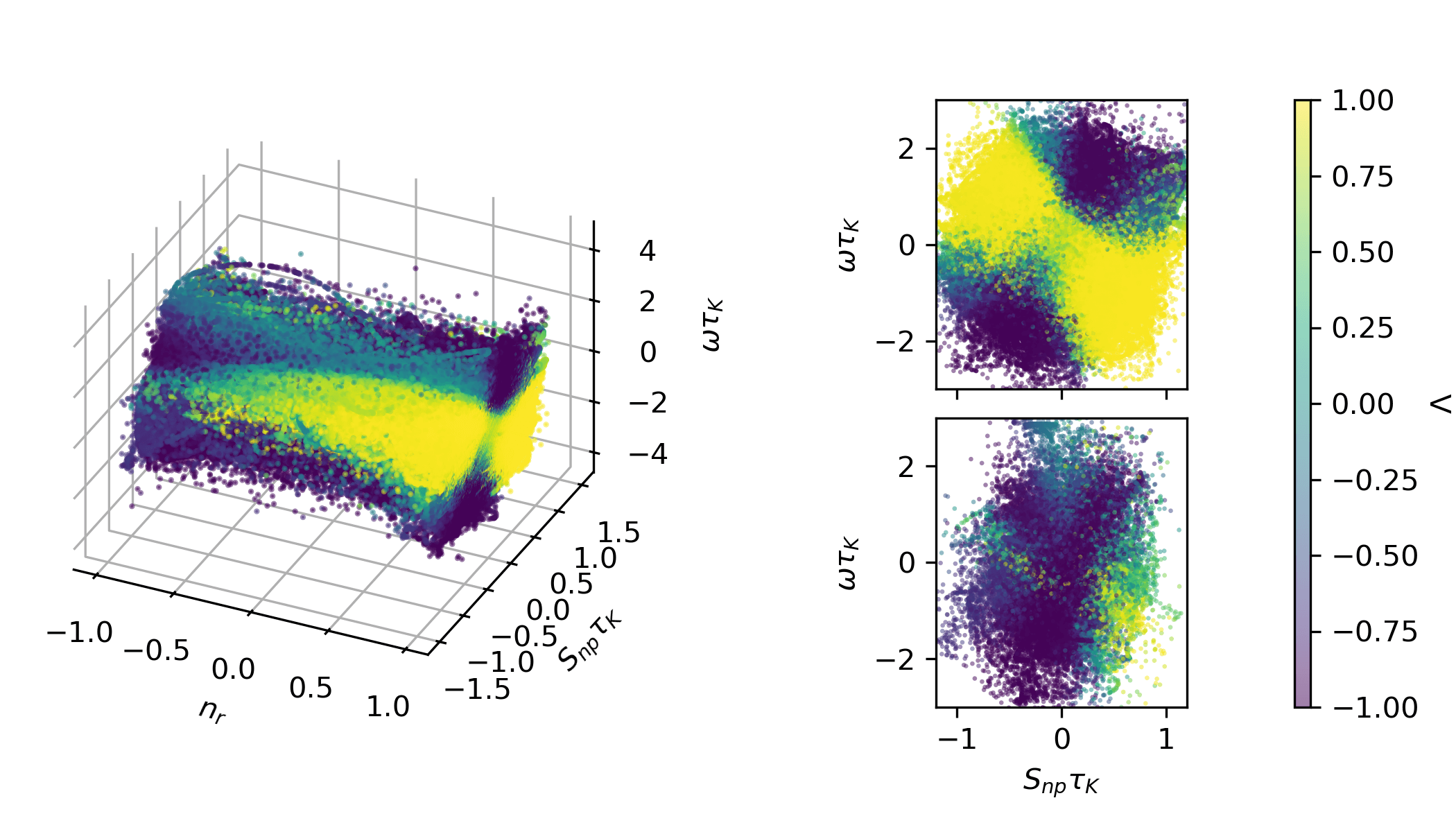}
    \put(5,37){(a)}
    \put(53,41){(b)}
    \put(53,21){(c)}        
    \end{overpic}
    \caption{Shape factor $\Lambda$ (color) for the smart strategy at ${\rm Ku}=10^7$ as a function of (a) $n_r$, $S_{np}$ and $\omega$; (b) $S_{np}$ and $\omega$ conditioned on $n_r\ge 0.8$; (c) $S_{np}$ and $\omega$ with $n_r\le -0.8$.
    Data is based on $4\times10^4$ trajectories.}
    \label{fig:kuInf}
\end{figure}

In the limit of large Kubo numbers, the smart strategy differs qualitatively from STO and performs substantially better. 
To clarify the mechanism, we analyzed correlations between the learned control $\Lambda$ and all three-signal combinations, identifying $n_r$, $S_{np}$, and $\omega$ as the dominant contributors to the policy. 
Figure~\ref{fig:kuInf}(a) shows $\Lambda$ as a function of these variables, revealing nontrivial correlations between $\lambda$ and strain/vorticity that depend sensitively on swimmer orientation. For orientations predominantly away from the initial position ($n_r>0$), the policy organises clearly in the $(S_{np},\omega)$ plane [Fig.~\ref{fig:kuInf}(b)]: $\Lambda>0$ when $S_{np}$ and $\omega$ have opposite signs or are both small, and $\Lambda<0$ when they share the same sign.
Thus, the smart strategy selects $\Lambda$ so that the strain-induced rotation term $\Lambda \ve p \cdot \ma S \cdot \ve n$ counteracts the vorticity contribution in the rotational dynamics~(\ref{eq:thetaDot}), reducing the net angular velocity. The swimmer thereby stabilises its orientation as it moves away from the initial position, thereby weakening the effect of persistent velocity gradients. This mechanism is absent in short-time optimal strategies and becomes effective only when the flow has long enough memory.

When the microswimmer is oriented towards the initial location ($n_r<0$), a different behaviour emerges: the smart policy strongly favours oblate shapes ($\Lambda<0$), largely independent of the instantaneous values of $S_{np}$ and $\omega$ [Fig.~\ref{fig:kuInf}(c)]. Oblate microswimmers preferentially align against the local flow direction and exhibit self-trapping~\citep{jeffery1922,borgnino2019Alignment}. Here, this mechanism acts as a protective response: by reducing its mobility, the microswimmer suppresses backward motion and avoids being advected back towards the initial position.

In addition to the two dominant mechanisms at large $\rm Ku$ identified in Fig.~\ref{fig:kuInf}(b, c), we identify a subdominant mechanism involving the second strain component, $S_{nn}$. In the smart strategy, $\Lambda$  is weakly negatively correlated with $S_{nn}$: extreme positive (negative) values of $S_{nn}$ are associated with negative (positive) $\Lambda$. For example, among the 1\% most negative values of $S_{nn}$, the probability of $\Lambda>0$ is 82\%. Because such events are rare, their contribution to the overall performance is limited. Still, the trend is consistent with prolate and oblate particles known to align with stretching and compressive directions in turbulence~\citep{pumir2011orientation,parsa2012rotation,gustavsson2014tumbling,zhan2014Accumulation,ni2014Alignment,zhao2016Why}. By actively modulating its shape, the smart swimmer disrupts this passive alignment, helping it avoid strongly stretching or compressive regions that would otherwise destabilise its orientation during navigation.

\subsection{Intermediate \texorpdfstring{$\rm Ku$}{Ku} transition}

\begin{figure}
    \centering
    \newlength{\RightFigHeight}
    \setlength{\RightFigHeight}{0.2\textheight}
    \newlength{\VSpace}
    \setlength{\VSpace}{0.5em}
    \begin{overpic}[width=1\textwidth,percent]{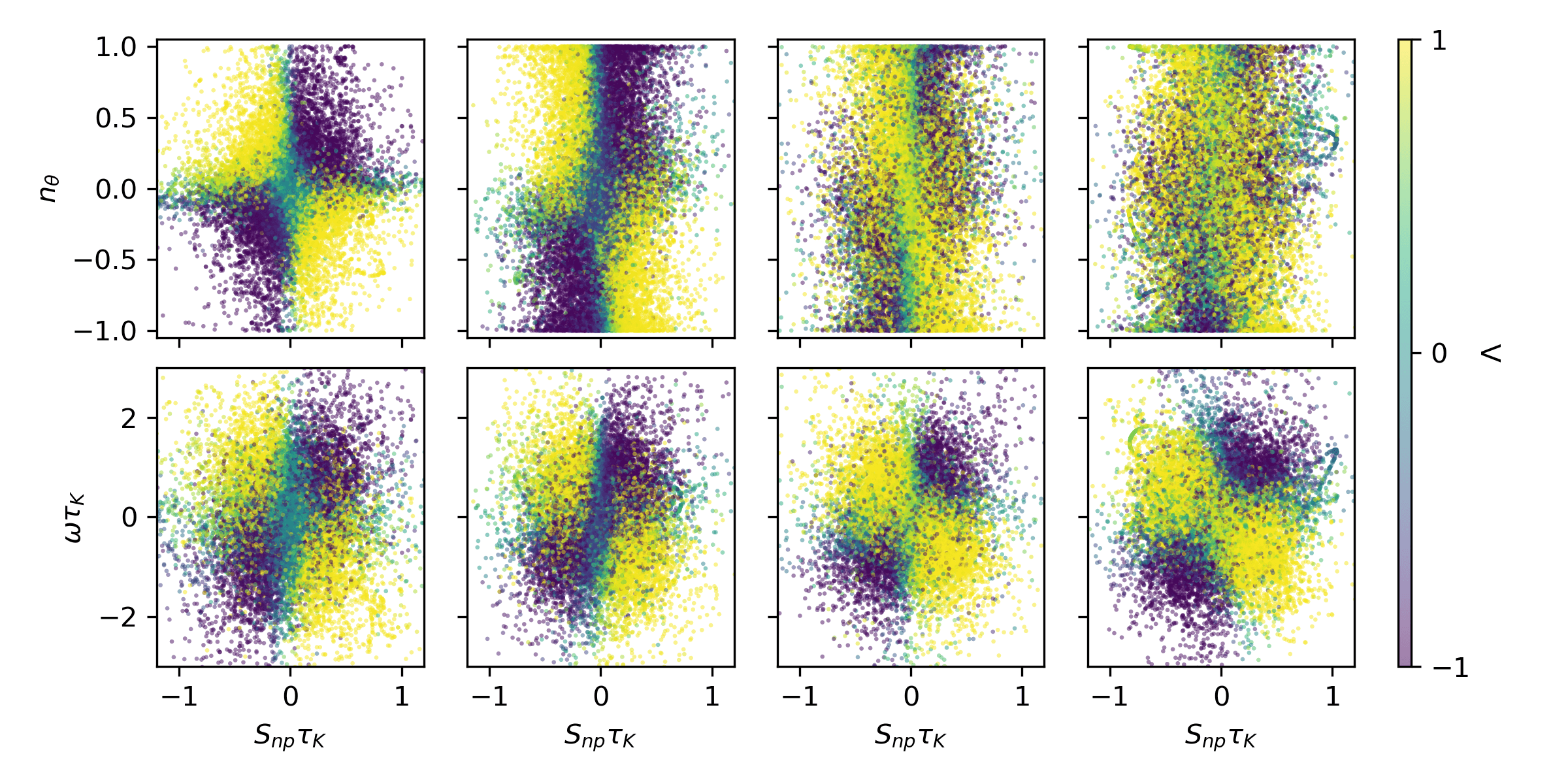}
        \put(11,45){(a)}
        \put(31,45){(b)}
        \put(51,45){(c)}
        \put(71,45){(d)}
        \put(11,24){(e)}
        \put(31,24){(f)}
        \put(51,24){(g)}
        \put(71,24){(h)}
    \end{overpic}
    \caption{Shape factor $\Lambda$ (color) for the smart strategy as a function of $S_{np}$ and $n_\theta$ (top row), and $S_{np}$ and $\Omega$ (bottom row). (a, e) ${\rm Ku}=0.1$. (b, f) ${\rm Ku}=1$. (c, g) ${\rm Ku}=10$. (d, h) ${\rm Ku}=10^7$. The data are based on $10^3$ trajectories.}
    \label{fig:ku_pattern}
\end{figure}
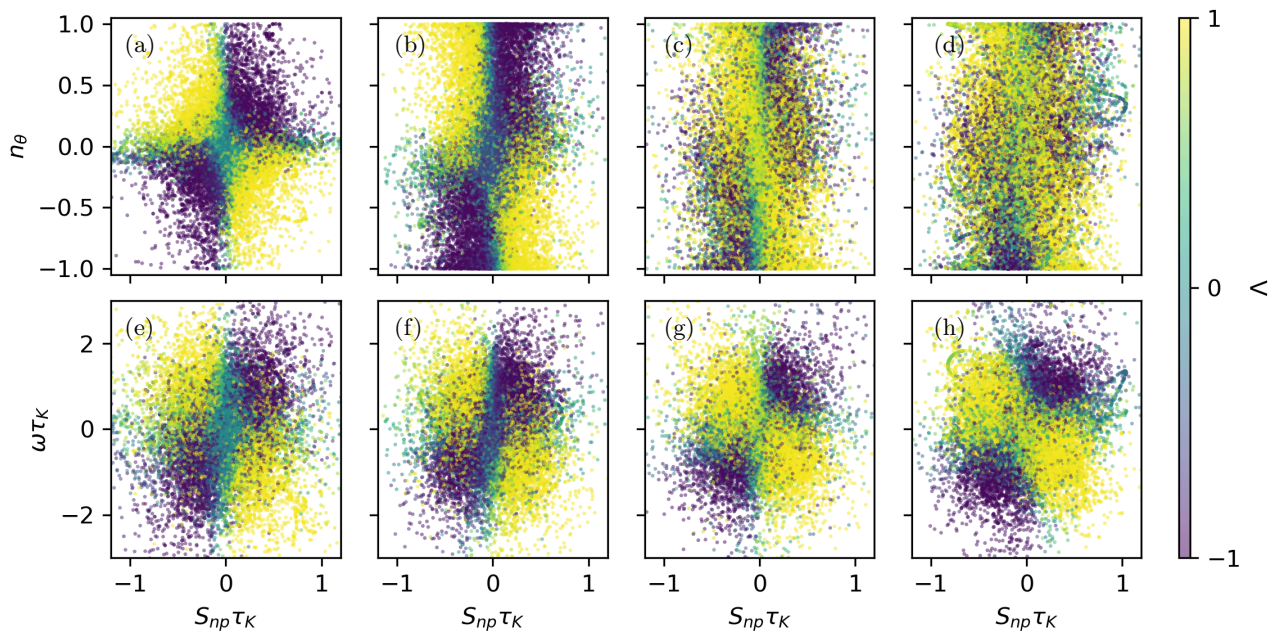

At intermediate values of the Kubo number, ${\rm Ku}=1$ and $10$, the smart strategies smoothly interpolate between the small-${\rm Ku}$ and large-${\rm Ku}$ limits. This gradual transition is evident in Fig.~\ref{fig:ku_pattern}, which shows the smart control action $\Lambda$ as a function of the relevant state variables for different values of ${\rm Ku}$.

At ${\rm Ku}=0.1$, the control landscape of $S_{np}$ and $n_\theta$ shows the characteristic dependence associated with the STO strategy (\ref{eq:sto}) [Fig.~\ref{fig:ku0.1}(b) and Fig.~\ref{fig:ku_pattern}(a)]. As ${\rm Ku}$ increases, this structure progressively weakens and eventually disappears [Fig.~\ref{fig:ku_pattern}(b–d)], reflecting that short-time optimisation becomes less effective in temporally correlated flows. Accordingly, the smart strategy relies less on this mechanism when the flow decorrelates more slowly.

A complementary trend is observed in the stabilisation mechanism of the quasi-steady regime. The $S_{np}$-$\omega$ correlation, which dominates the smart strategy at ${\rm Ku}=10^7$ [Fig.~\ref{fig:kuInf}(b) and Fig.~\ref{fig:ku_pattern}(h)], transitions gradually as ${\rm Ku}$ decreases [Fig.~\ref{fig:ku_pattern}(e–h)]. 
While this correlation persists even at small Kubo numbers, its physical origin changes. At ${\rm Ku}=0.1$, the correlation emerges because the vorticity acts as a perturbation to the orientation $n_\theta$ (see section~\ref{sec:results:smallKu}), which correlates $\omega$ with $n_\theta$. Hence $\omega$ acquires a correlation with $S_{np}$ via the $n_\theta$-$S_{np}$ correlation in the STO strategy~(\ref{eq:sto}).
As ${\rm Ku}$ increases, long-time orientation stabilisation becomes more important as flow correlation time increases.

Remarkably, because the strategies are trained independently at each value of ${\rm Ku}$, the observed smooth evolution of control patterns is not imposed by construction but arises naturally from the interaction between the microswimmer and the flow. This continuity suggests that efficient microswimmer navigation is governed by a small set of underlying physical mechanisms whose relative importance is controlled by the flow regime. The smart microswimmer adapts by continuously reweighting these mechanisms in response to the temporal and spatial structure of the flow, rather than switching abruptly between qualitatively distinct strategies.

\subsection{From learned smart strategies to a minimal analytical model of navigation}

\begin{figure}
    \centering
    \begin{overpic}[width=0.95\textwidth,percent]{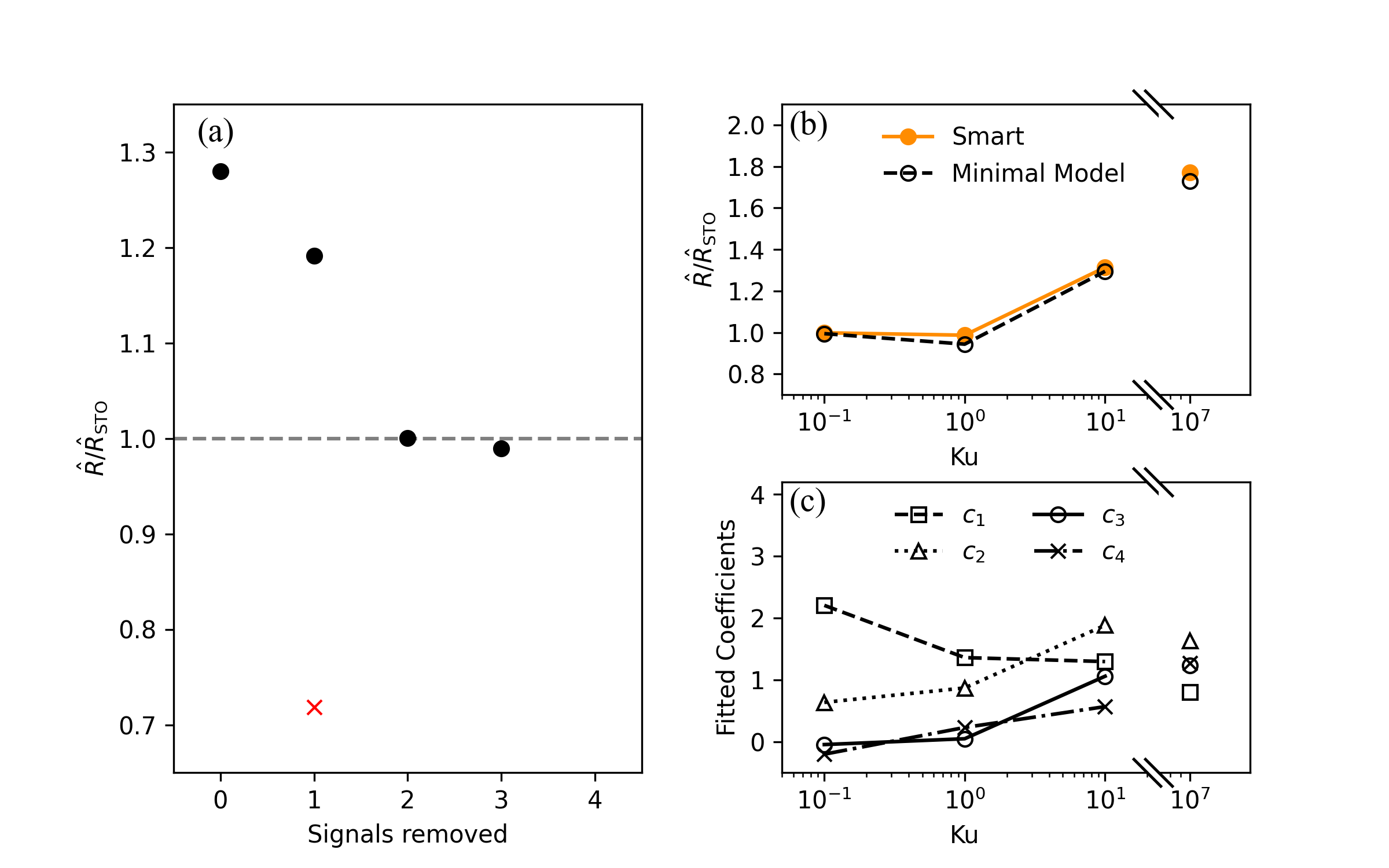}
        \put(18,49.5){\small$(n_r, n_\theta, S_{nn}, S_{np}, \omega)$}
        \put(25,43.2){\small$(n_r, n_\theta, S_{np}, \omega)$}
        \put(15,33){\small$(n_r, n_\theta, S_{np})$}
        \put(35,33){\small$(n_\theta, S_{np})$}
        \put(13,14){\small$(n_r, n_\theta, S_{nn}, \omega)$}
        \end{overpic}
        \caption{(a) Relative performance of smart strategies compared to the STO baseline at ${\rm Ku}=10$, trained with different sets of input signals. Brackets denote the signals available to each strategy. (b) Performance as a function of ${\rm Ku}$ for the smart strategy and the minimal model~(\ref{eq:fit}). (c) Fitted coefficients of the minimal model as functions of ${\rm Ku}$.}
    \label{fig:ablation}
\end{figure}
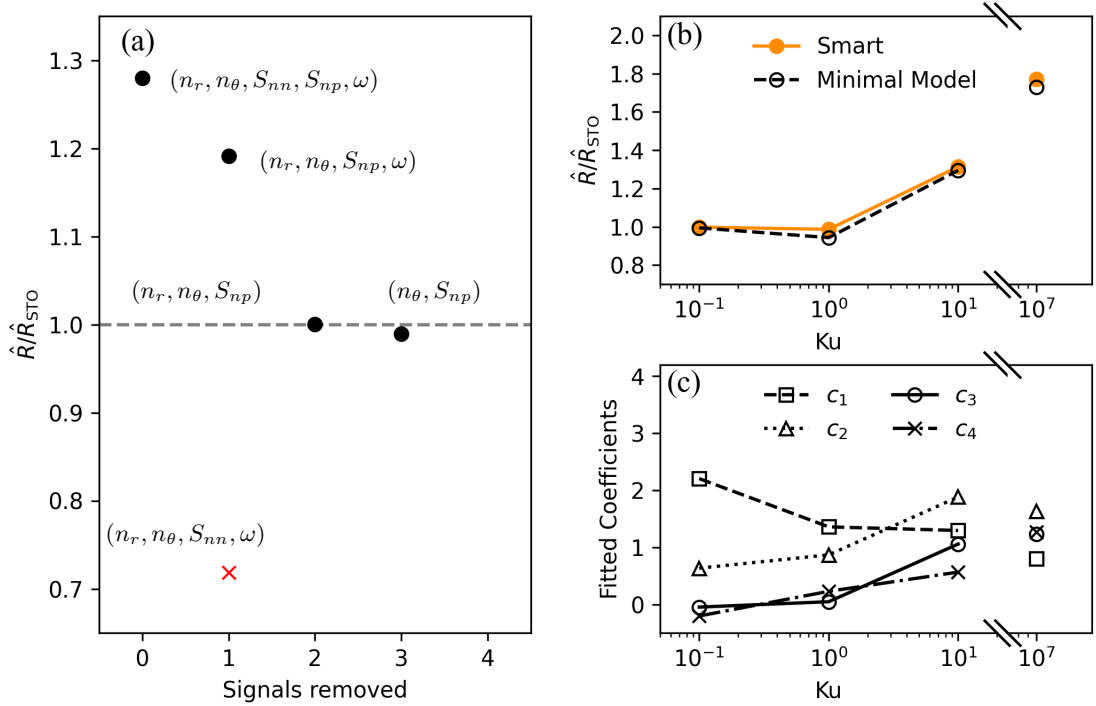

In Sections~\ref{sec:results:smallKu} and \ref{sec:results:largeKu}, we identified four distinct physical mechanisms that underpin efficient navigation in stochastic flows:
(i) the short-time optimization (STO) mechanism, which relies on $S_{np}$ and $n_\theta$ and is effective at small ${\rm Ku}$;
(ii) an orientation-stabilisation mechanism, controlled by $S_{np}$ and $\omega$, which dominates at large ${\rm Ku}$;
(iii) a self-trapping mechanism, associated with negative radial orientation $n_r<0$, also relevant at large ${\rm Ku}$; and
(iv) an extreme strain-avoidance mechanism, governed by $S_{nn}$, which contributes primarily in the large-${\rm Ku}$ regime.

To quantify the relative importance of these mechanisms in the smart strategy, we conduct a ``signal-ablation" experiment at ${\rm Ku}=10$. In this procedure, the microswimmer is trained multiple times using progressively reduced sets of input signals. By selectively removing the signals associated with specific mechanisms, the corresponding behaviours are suppressed, allowing a direct assessment of their relevance to strategy performance.

Figure~\ref{fig:ablation}(a) reports the performance of the trained strategies relative to the STO baseline. When all signals are available, the smart strategy outperforms the STO baseline by nearly $30\%$. Removing $S_{nn}$ suppresses the extreme strain-avoidance mechanism and leads only to a marginal decrease in performance, consistent with its secondary role (see Section~\ref{sec:results:largeKu}). Further removing $\omega$, such that only $n_r$, $n_\theta$, and $S_{np}$ remain, eliminates the orientation-stabilisation mechanism and causes an abrupt drop in performance to the level of the STO strategy. Excluding $n_r$ as well results in just a mild decrease, since the remaining signals $n_\theta$ and $S_{np}$ are sufficient to recover the STO mechanism. Beyond Fig.~\ref{fig:ablation}(a), any policy trained on three or fewer signals is outperformed by STO (not shown).

Amongst all inputs, $S_{np}$ emerges as the most critical signal. This is expected because it underlies both the STO and orientation-stabilisation mechanisms, and this is confirmed by the red cross in Fig.~\ref{fig:ablation}(a): removing $S_{np}$ while retaining all other signals leads to a failure of the smart strategy.

Guided by this mechanistic understanding, we propose a minimal analytical model for the smart strategy,
\begin{equation}
\begin{split}
    \Lambda = \Lambda_{\rm max} \tanh \Bigl[ & -c_1 n_\theta\, {\rm sign}(S_{np}) - c_2 H(n_r)\omega \tau_{\rm K}\, {\rm sign}(S_{np}) \\
    & + c_3 n_r - c_4 \tanh(S_{nn} \tau_{\rm K}) \Bigr]\,,
    \label{eq:fit}
\end{split}
\end{equation}
where $H(x)$ denotes the Heaviside function and $c_1$--$c_4$ are ${\rm Ku}$-dependent coefficients obtained by fitting the model to the learned smart strategies for different ${\rm Ku}$ using the curve\_fit function from the SciPy library \citep{2020SciPy-NMeth}. The model combines the four identified mechanisms through a weighted sum, while the outer hyperbolic tangent constrains the control output to the admissible range of $\Lambda$.

Each term in Eq.~(\ref{eq:fit}) has a clear physical interpretation. The first term represents the STO strategy. The second term corresponds to the orientation-stabilisation mechanism and is activated only when $n_r>0$ via the Heaviside function. The third term captures the negative correlation between $\Lambda$ and $n_r$ responsible for the self-trapping mechanism (see Fig.~\ref{fig:kuInf}), while the final term models the strain-avoidance response associated with $S_{nn}$.

The performance of the minimal model is shown in Fig.~\ref{fig:ablation}(b). Its efficiency is nearly indistinguishable from that of the full smart strategy across the range of considered ${\rm Ku}$ values, demonstrating that the model captures the essential physics underlying efficient navigation. 

Beyond reproducing performance, the minimal model provides direct insight into how navigation strategies evolve across flow regimes, as the fitted coefficients quantify the relative weight of each mechanism in the composite strategy. Figure~\ref{fig:ablation}(c) shows the coefficients as functions of ${\rm Ku}$. Consistent with the STO mechanism being effective only at small ${\rm Ku}$, the magnitude of $c_1$ decreases monotonically with increasing ${\rm Ku}$. In contrast, the magnitudes of the coefficients associated with large-${\rm Ku}$ mechanisms ($c_2$–$c_4$) tend to increase with ${\rm Ku}$.

\subsection{Transfer learning in DNS turbulent flows}

\begin{figure}
    \centering
    \includegraphics[width=0.95\linewidth]{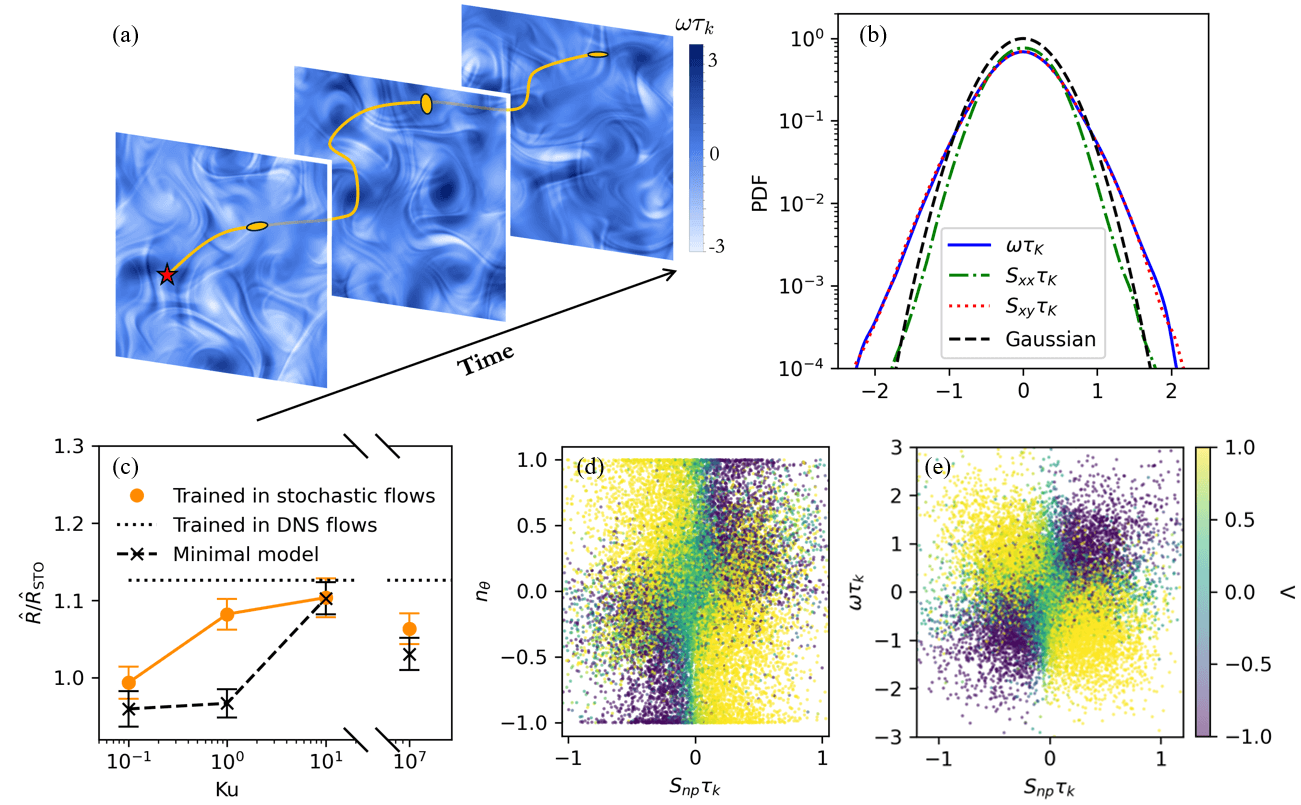}
    \caption{(a) Temporal snapshots of the vorticity field of the turbulent flow obtained from DNS. The orange curve is a qualitative illustration of a microswimmer trajectory. The red star indicates its initial position, and the orange spheroids represent its shape and position in the flow field at successive time snapshots.
    (b) Probability distribution of vorticity $\omega$ (blue),  normal strain $S_{xx}$ (green), and shear strain $S_{xy}$ (red) of the turbulent flow, compared with a Gaussian distribution (black).
    (c) Performance of the strategies relative to the STO baseline [Eq.~(\ref{eq:sto})] in two-dimensional turbulence. The orange curve represents the strategy learned in stochastic flows at different ${\rm Ku}$. The black dashed curve corresponds to the minimal model~(\ref{eq:fit}) using the coefficients from Fig.~\ref{fig:ablation}(c). The black dotted line shows the performance of the strategy trained in the turbulent flow. Error bars indicate standard deviations over random initial positions and independent flow realizations. 
    (d,e) Shape factor $\Lambda$ projected onto (d) $S_{np}$ and $n_\theta$; (e) $S_{np}$ and $\omega$ for the DNS-trained strategy.}
    \label{fig:dnsFlow}
\end{figure}

We test the robustness of the smart strategies by transferring them to two-dimensional turbulence generated by DNS. The flow consist of an incompressible velocity field, $\bm \nabla \cdot \bm u = 0$, governed by the Navier–Stokes equations
\begin{equation}
    \partial_t \ve u+\ve u\cdot\ve \nabla \ve u=-\nabla p +\nu\Delta \ve u-\alpha \ve u +\ve F\,, 
    \label{eq:NSE}
\end{equation}    
where $p$ is pressure, $\nu$ the viscosity, $\ve F$ a large-scale external forcing that drives an enstrophy cascade at smaller scales~\citep{boffetta2012}, and the term $-\alpha \ve u$ represents friction preventing energy accumulation at large scales. Equation~(\ref{eq:NSE}) is integrated using a standard pseudospectral method (see Appendix~\ref{app:simulations}).
Representative snapshots of the flow vorticity field and the statistical properties of flow velocity gradients are shown in Figs.~\ref{fig:dnsFlow}(a,b).

The orange curve in Fig.~\ref{fig:dnsFlow}(c) shows the DNS performance of the smart strategies trained in stochastic flows at different ${\rm Ku}$. 
Performance depends on ${\rm Ku}$: the strategy trained at ${\rm Ku}=10$ performs best, exceeding the STO baseline and nearly matching a strategy trained in DNS turbulence (black dotted line). 
This indicates that stochastic flows at ${\rm Ku}=10$ capture key dynamical features of DNS turbulence, consistent with earlier observations that large but finite-${\rm Ku}$ models reproduce essential Lagrangian statistics of turbulent transport of passive and active particles~\citep{gustavsson2016Preferential,borgnino2019Alignment,bec2024Statistical}.

Figures~\ref{fig:dnsFlow}(d, e) show that the $(S_{np}, n_\theta)$ and $(S_{np}, \omega)$ correlations underlying the STO and orientation-stabilisation mechanisms in stochastic flows, persist in the DNS-trained strategy. Moreover, the close agreement between the minimal model~(\ref{eq:fit}) [black dashed in Fig.~\ref{fig:dnsFlow}(c)] and the transferred strategy at ${\rm Ku}=10$ further confirms that these mechanisms capture the essential physics of efficient navigation and remain robust in realistic turbulence.

Residual differences between the DNS-trained and transferred strategies likely reflect fine-scale structural discrepancies between stochastic and Navier-Stokes turbulence. Away from ${\rm Ku}=10$, these mismatches grow and transfer performance degrades. An exception is ${\rm Ku}=1$, where the transferred strategy performs unexpectedly well, outperforming both STO and the minimal model. 
This suggests that at intermediate ${\rm Ku}$ the smart strategy effectively combines short-time optimisation with long-time stabilisation, enhancing robustness to the broader range of scales present in DNS turbulence.

\section{Conclusions}
\label{sec:conclusions}

We have investigated the navigation of shape-changing spheroidal microswimmers tasked with maximising displacement from their initial position. Using reinforcement learning, we identified smart strategies in which the microswimmer adaptively adjusts its shape factor, $\Lambda$, in response to local orientation and fluid-gradient signals.

Across flows with different temporal correlations, characterised by the Kubo number ${\rm Ku}$, defined as the ratio of flow correlation to advection timescales, the learned smart strategies consistently outperform both a naive fixed-shape baseline and the short-time optimisation baseline~(\ref{eq:sto}). Efficient navigation relies upon multiple complementary mechanisms: the short-time optimisation mechanism, based on $n_\theta$ and $S_{np}$, maximises short-time displacement; orientation stabilisation, driven by $S_{np}$ and vorticity $\omega$, reduces rotation induced by fluid gradients and stabilises the orientation when pointing away from the initial position ($n_r>0$);
adopting an oblate shape when pointing toward the initial location ($n_r<0$) prevents backward motion;  and avoidance of strong normal strain ($S_{nn}$) prevents destabilising alignment. The relative contribution of these mechanisms varies with the flow’s correlation time, as quantified by the minimal model~(\ref{eq:fit}).

The universality of these mechanisms is confirmed by transferring strategies trained in stochastic flows to fully resolved DNS turbulence. Strategies learned at ${\rm Ku}=10$ perform nearly as well in turbulence as those trained directly in the DNS environment, showing that the stochastic flow model effectively captures essential Lagrangian dynamics and provides a cost-effective framework for exploring navigation strategies and their underlying physics.
Beyond quantitative performance, these results highlight a novel navigation paradigm in which adaptive morphology, rather than direct actuation, enables microswimmers to exploit local mechanical cues.

Future work could explore more complex tasks, such as point-to-point navigation, incorporate energetic costs of active shape control, which may reveal qualitatively different strategies~\citep{piro2024Energetica}, and investigate hydrodynamic or mechanical interactions among multiple microswimmers, potentially enabling cooperative or adversarial strategies in complex flows~\citep{borra2022}.
Another future direction is to extend the navigation strategy to three-dimensional flows. The additional degrees of freedom increase the complexity of the state space and dynamics, presenting challenges for the control strategy.
Furthermore, while this study employs an idealized spheroidal model, future research may explore alternative adaptive morphologies, such as swimmers with flexible~\citep{tarama2014Deformable} or composite~\citep{zou2022Gait} body structures.

\section*{Acknowledgements}
    The authors thank Antonio Celani and Navid Mousavi for helpful discussions.
    We acknowledge financial support from Vetenskapsrådet, Grants No. 2018-03974, No. 2023-03617 (J.Q. and K.G.), and No. 2021-4452 (B.M.). K.G., B.M., and J.Q. acknowledge support from the Knut and Alice Wallenberg Foundation, Grant No. 2019.0079. This work was also supported by the European Research Council (ERC) under the European Union’s Horizon 2020 research and innovation programme (Grant Agreement No. 882340).

\appendix

\begin{appendix}

\section{Numerical simulation}
\label{app:simulations}

We compute microswimmer trajectories using the Eulerian–Lagrangian scheme of~\cite{qiu2022Navigation,qiu2022Active} and~\cite{mousavi2024Efficient,mousavi2025Short}. In this scheme, Eq.~(\ref{eq:eom}) is integrated with a second-order method with time step $\delta t$. At each step, the fluid velocity and its gradients at the swimmer position are obtained by second-order interpolation from the Eulerian flow field. 

The two-dimensional stochastic flow is stored on a uniform $400\times400$ Eulerian grid with spacing $0.025\ell_{\rm f}$. As explained in Section~\ref{sec:flowmodel}, velocity field in Eq.~(\ref{eq:flowmodel}) is generated from $M=5$ stream-function components with coefficients evolving via an Ornstein-Uhlenbeck process using the same time step $\delta t$. 
To allow unbounded swimmer motion, the domain is periodically extended in both spatial directions. Our results are insensitive to the domain size, because the flow and its gradients decorrelate before a swimmer traverses the domain. This was confirmed using a ten times larger domain.

Two-dimensional turbulence is obtained by direct numerical simulations of Eq.~(\ref{eq:NSE}) by means of a standard 2/3 de-aliased pseudospectral method on a bi-periodic square domain of size $2\pi$ using $512\times512$ grid points. The flow is forced at large scales by a Gaussian, temporally uncorrelated forcing $\ve F$ peaked at wavenumber $k_{\rm f}=3$, in order to induce a direct enstrophy cascade. Friction $-\alpha \bm v$, preventing energy buildup at large scales, is numerically implemented via a hypo-friction term~\citep{boffetta2012}.
We stored the velocity and gradient fields at about every tenth of the Kolmogorov characteristic time.
During the simulation of microswimmers, linear interpolation is used to obtain snapshots between successive stored flow fields. 
The flow velocity and its gradients are then interpolated to the position of microswimmers using the same second-order Lagrangian scheme as in the stochastic flow model. Similar to the stochastic flow, the turbulent flow is also periodically extended, and the domain size is sufficiently large.

Velocity and gradient fields are stored at intervals of approximately one-tenth of the Kolmogorov time. During swimmer simulations, intermediate flow snapshots are obtained by linear interpolation in time, and spatial interpolation to swimmer positions uses the same second-order Lagrangian scheme as in the stochastic model. The turbulent flow is also periodically extended, and the domain is sufficiently large to avoid finite-size effects.

In both the stochastic model and DNS turbulence, the microswimmer dynamics is integrated with the time step $\delta t=0.01\tau_{\rm K}$. This choice resolves both flow fluctuations and swimmer dynamics, even for the most rapidly evolving case (${\rm Ku}=0.1$), where the flow correlation time $\tau_{\rm f}$ is still an order of magnitude larger than $\delta t$.

\section{Reinforcement learning algorithm}\label{app:RL}

We use Proximal Policy Optimisation (PPO)~\citep{schulman2017Proximal} to identify smart deformation strategies for a microswimmer tasked with maximising its displacement from its initial position. PPO is chosen for its stability in noisy environments and its robust performance in continuous action spaces. It is an on-policy actor–critic reinforcement learning algorithm~\citep{sutton2018Reinforcementa} that optimises a stochastic policy $\pi(\ve a_i|\ve s_i)$, defined as the probability distribution of taking action $\ve a_i$ given state $\ve s_i$ at time $t=iT_{\rm u}$.

The objective is to approximate the optimal policy that maximises the expected discounted return, $R=\sum_{i=0}^{N} \gamma^{i}r_i$,
where the discount factor $\gamma\in(0,1)$ balances future and instantaneous rewards. 
PPO achieves this by maximising the clipped surrogate objective function 
\begin{equation}
    J(\ve w) = \mathbb{E}\left\{
    {\rm min}[P_r(\ve w) {A}_i\,,{\rm clip}(P_r(\ve w), 1-\epsilon, 1+\epsilon)A_i]
    \right\}\,,
\end{equation}
where $\ve w$ denotes the parameters of the neural-network, $\mathbb{E}$ is the expectation over the sampled trajectories and $P_r(\ve w)=\pi_{\ve w}(\ve a_i|\ve s_i)/ \pi_{\ve w_{old}}(\ve a_i|\ve s_i)$ is the probability ratio between the updated and previous policies.
The advantage $A_i$ measures how much the selected action outperforms the expected action under the current policy, which is computed using Generalised Advantage Estimation following the standard implementation~\citep{schulman2017Proximal}.
Maximising $P_r(\ve w)A_i$ increases the likelihood of favourable actions, while the clipping and minimising operations restrict excessively large policy updates, thus improving training stability~\citep{schulman2017Proximal}.
For instance, when an action is favourable ($A_i>0$), the optimisation tends to increase $P_r$. With the clipping and minimisation, however, $J(\ve w)$ is capped at $(1+\epsilon)A_i$, preventing $P_r$ from growing arbitrarily large. 
Conversely, when an action is unfavoured ($A_i<0$), the optimisation decreases $P_r$. If $P_r<1-\epsilon$, $J(\ve w)$ is limited to $(1-\epsilon)A_i$, preventing an excessive reduction of $P_r$.

PPO uses two neural networks: an Actor Network and a Critic Network. The Actor Network takes the state $\ve s_i$ as input and outputs the policy $\pi(\ve a_i|\ve s_i)$. We assume a Gaussian action distribution, with the network outputting the mean and the standard deviation represented by a trainable scalar parameter. During training, actions are sampled from this distribution and then passed through a $\tanh$ function to bound the action values. Since this transformation changes the action distribution, the corresponding probability ratio and entropy terms are modified accordingly in the objective function $J(\ve w)$. During evaluation, actions are sampled in the same way except that the standard deviation is set to zero, such that the actions are deterministically equal to the mean of the distribution.
The Critic Network also takes $\ve s_i$ as input and outputs a scalar state-value estimate $V(\ve s_i)$, which is used in the calculation of the advantage function,
\begin{equation}
\begin{aligned}
    &A_i = \delta_i + (\gamma\lambda_{\rm GAE})\delta_{i+1}+\dots +(\gamma\lambda_{\rm GAE})^{N-i+1}\delta_{N-1}\,, \\
    &\text{where}~~\delta_i=r_i+\gamma V(\ve s_{i+1})-V(\ve s_i)\,,
\end{aligned}
\end{equation}
and $\lambda_{\rm GAE}$ is the Generalised Advantage Estimation parameter (see~\cite{schulman2017Proximal} for details).

Both the Actor and Critic Networks are fully connected feed-forward neural networks with two hidden layers of 64 neurons each, using $\tanh$ activation functions. For the Actor Network, a linear output layer is used for the mean of the Gaussian action distribution. The Critic Network employs a linear output layer for the value estimate.
The network parameters are updated via gradient ascent on $\nabla_{\ve w} J(\ve w)$.

The main hyperparameters used in PPO training are summarised as follows. The discount rate is set to $\gamma=0.99$, the PPO clipping parameter is set to $\epsilon=0.1$, and the Generalised Advantage Estimation parameter is set to $\lambda_{\rm GAE}=0.99$. 
An entropy regularisation term with coefficient 0.005 is included to promote exploration and prevent early convergence of policy.
The Actor and Critic Networks are optimised using the Adam optimiser as implemented in PyTorch~\citep{paszke2019PyTorch}. The learning rates decrease linearly from $4\times10^{-4}$ for the Actor Network and $8\times10^{-5}$ for the Critic Network to zero over 5000 training episodes. Each policy update comprises four epochs, each over 200 minibatches with a batch size of 25.

The reinforcement learning algorithm is coupled to the numerical simulation of microswimmer dynamics. In each episode, ten microswimmers are initialised with random initial positions and orientations, as described in Section~\ref{sec:methods:rl}, and simulated in parallel to collect trajectory data efficiently. At each decision time $t=iT_{\rm u}$, the state variables are computed, and the PPO agent outputs actions that set the shape factors of the microswimmers for the subsequent time interval. The microswimmer trajectories are then advanced in time using the numerical scheme described in Appendix~\ref{app:simulations}. This online coupling allows the agent to learn directly from physically resolved Lagrangian dynamics.

\begin{figure}
    \centering
    \includegraphics[width=0.5\linewidth]{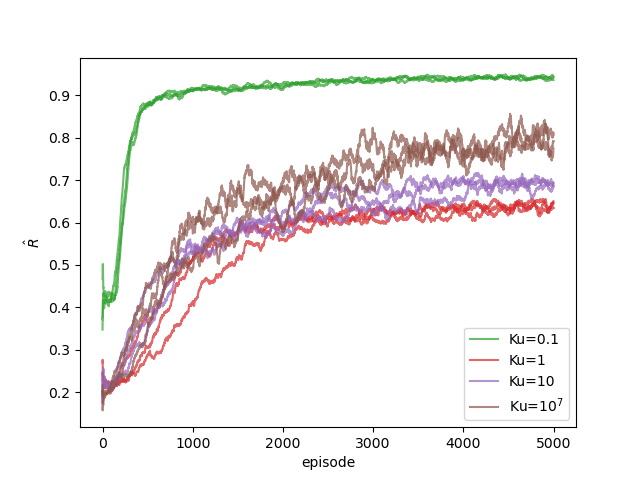}
    \caption{Training curves of the performance index $\hat R$ against training episode. For each ${\rm Ku}$, three independent runs with different random seeds are shown.}
    \label{fig:trainingCurves}
\end{figure}

Training is performed for 5000 episodes, and each run uses three independent random seeds to ensure statistical robustness. Figure~\ref{fig:trainingCurves} shows the evolution of the navigation performance as a function of training episode. The performance saturates after a sufficient number of episodes, indicating convergence of the learned strategy. The resulting strategies after training are evaluated on flow realisations that differ from the training data to test their generalizability. All results shown in the paper are obtained from this evaluation process.

\section{Robustness against rotational noise and other parameters}\label{app:robustness}

\begin{figure}
    \centering
    \begin{overpic}[width=0.8\textwidth]{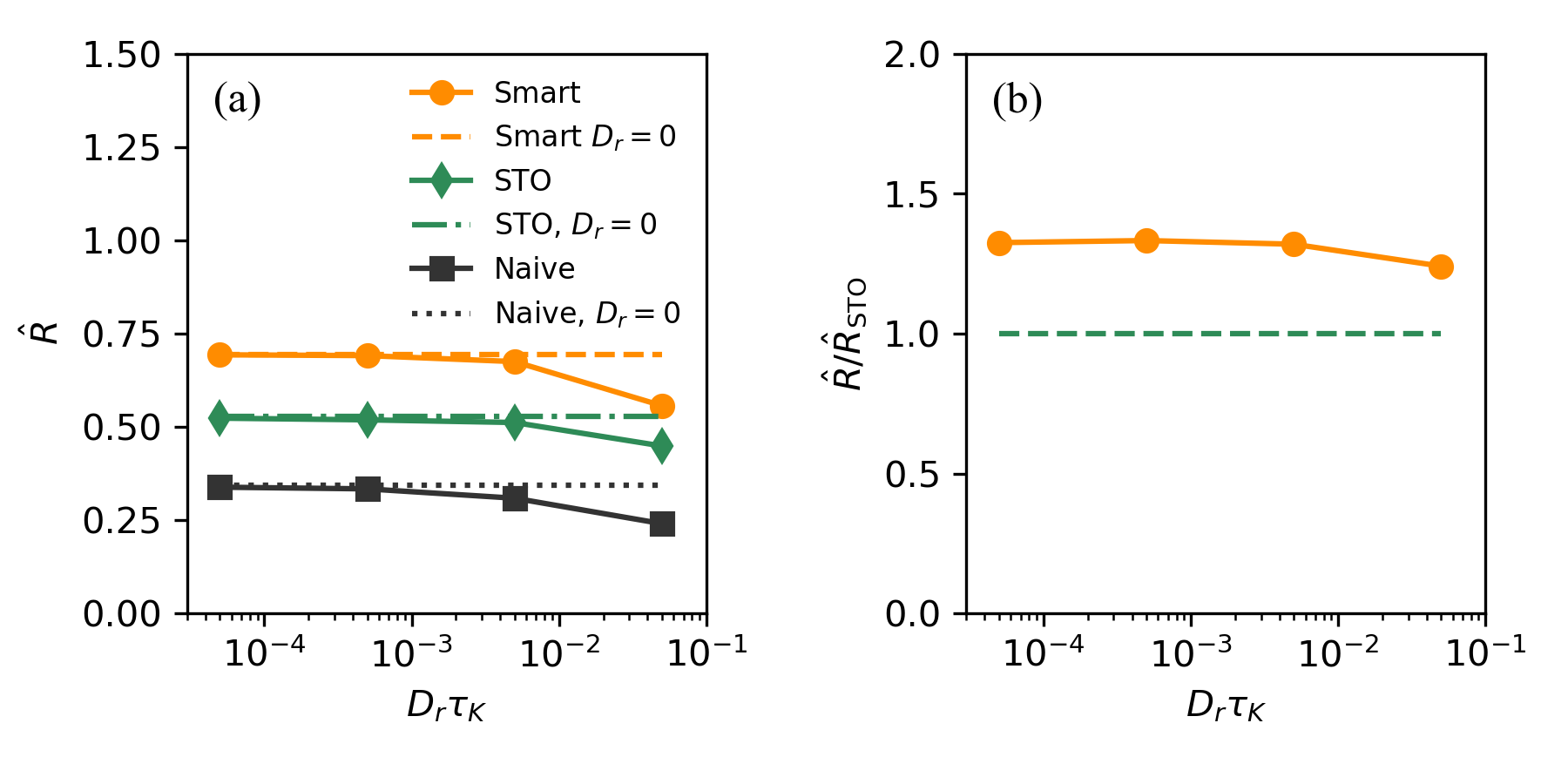}
    \end{overpic}
    \caption{Robustness against rotational noise. (a) Performances of smart (orange), naive (black), and STO (green) strategies at different noise levels. The smart strategy is trained at ${\rm Ku}=10$, $D_{\rm r}=0$. (b) Performance of the smart strategy relative to the STO baseline across varying noise levels.}
    \label{fig:robustness-noise}
\end{figure}

Here, we test the robustness of the smart strategy learned at ${\rm Ku}=10$. 

First, we consider the robustness against rotational noise. The two-dimensional Brownian noise is implemented as a displacement in $\ve n$ with a variance of $\sqrt{2D_{\rm r} \delta t}$, where $D_{\rm r}$ is the rotational diffusion coefficient, and $\delta t$ is the time step of the integration of dynamics. 
Figure~\ref{fig:robustness-noise} shows the performance of the smart, naive, and STO strategies at different noise levels. 
The performance of the smart strategy begins to decrease when $D_{\rm r}\tau_{\rm K} > 5\times10^{-3}$, yet it still outperforms both baselines even at the largest noise level considered.

\begin{figure}
    \centering
    \includegraphics[width=\textwidth]{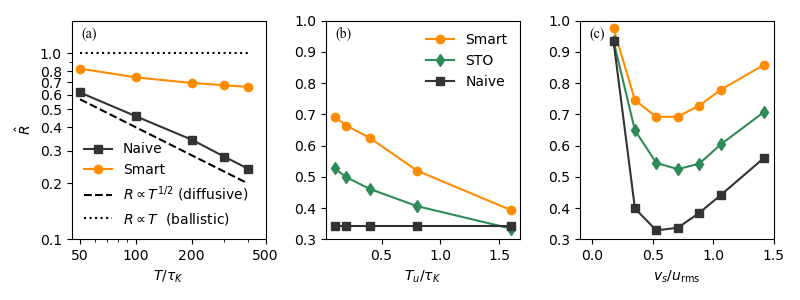}
    \caption{Performance index $\hat R$ (\ref{eq:reward2}) of smart (orange), naive (black), and STO (green) strategies evaluated for different (a) episode durations $T$; (b) update intervals $T_{\rm u}$; (c) swimming speeds $v_{\rm s}$. 
    The smart strategy was trained at ${\rm Ku}=10$, $T/\tau_{\rm K}=200$, $T_{\rm u}/\tau_{\rm K}=0.1$, and $v_{\rm s}/u_{\rm rms} = 0.707$.}
    \label{fig:robustness}
\end{figure}

Figure~\ref{fig:robustness}(a) shows the robustness of the smart strategy against different episode durations $T$. The smart strategy trained at $T=200\tau_{\rm K}$ performs well also at other episode durations, yielding an almost constant normalised performance $\hat{R}$. This indicates that, with the smart strategy (orange curve), the travel distance increases almost linearly with time $|\ve x(t_N)-\ve x(t_0)|\propto T$. Such ballistic-like transport surpasses the diffusive dispersion of naive microswimmers with constant shape whose normalised performance scales as $\hat{R}\propto T^{-1/2}$ [Fig.~\ref{fig:robustness}(a), black solid curve]. 
Figure~\ref{fig:robustness}(b) illustrates the sensitivity of the smart and STO strategies to the update interval $T_{\rm u}$. The performance of both strategies decreases as $T_{\rm u}$ increases, reflecting the growing delay between state measurement and action execution. Since both strategies rely on signals from the local velocity gradients, they fail when $T_{\rm u}$ becomes comparable to or exceeds the Kolmogorov timescale $\tau_{\rm K}$, which sets the correlation time of these gradients. Nevertheless, the smart strategy continues to outperform both the STO and naive baselines, whereas the STO strategy approaches the naive one when $T_{\rm u} \gtrsim \tau_{\rm K}$.

Finally, Fig.~\ref{fig:robustness}(c) demonstrates the robustness of strategies with respect to the swimming speed $v_{\rm s}$. The smart strategy outperforms both the STO and naive baselines on all the values of $v_{\rm s}$. Interestingly, deviating from the specific $v_{\rm s}$ used during training, either by increasing or decreasing the speed, results in increased $\hat{R}$. The dependence of $\hat{R}$ on $v_{\rm s}$ arises from the definition of normalisation $\hat{R}=R/R_{\rm max}^{(0)}$. In the limit as $v_{\rm s}\rightarrow0$, $\hat{R}$ tends to infinity because the dimensional travel distance $R$ is bounded by the flow advection (scaled by $u_{\rm rms}$), while the reference $R_{\rm max}^{(0)}=v_{\rm s}T$ vanishes. In contrast, in the limit of large $v_{\rm s}$, the influence of flow becomes negligible, and $\hat{R}$ is expected to converge to unity.
Overall, the superior performance of the smart strategy relative to the baselines indicates that its mechanisms are robust to variations in swimming speed. 

\section{Optimal control theory}
\label{app:OCT}

In this appendix, we briefly outline how optimal control theory leads to the escape strategy discussed in Sec.~\ref{sec:baselines}. We consider a spheroidal self–propelled microswimmer whose translational and orientational dynamics in two dimensions are governed by Eqs.~\eqref{eq:tran} and~\eqref{eq:thetaDot}, which we report here for convenience
\begin{align}
    \dot{\ve x} &= \ve u + v_{\rm s} \ve n\,,     \label{eq:motion1}    \\
    \dot{\theta}&=\frac{1}{2}\omega + \Lambda S_{np}\,, \label{eq:motion2}
\end{align}
where $\ve n = (\cos\theta,\sin\theta)$ and the shape factor for spheroids takes the simple form $\Lambda=\frac{\lambda^2-1}{\lambda^2+1}$, with $\lambda$ the spheroid aspect ratio.

We now aim at finding the optimal protocol for the control $\lambda$, or equivalently for the shape factor $\Lambda$, that maximises its distance from the initial position $\bm{x}(t_0)$ over a fixed episode duration $T$. This is formulated by minimising the cost functional~\eqref{eq:cost} and identifying the state vector with $\bm{q}=(\bm{x},\theta)$, where $\bm{x}$ denotes the swimmer position and $\theta$ its orientation.

To apply Pontryagin’s minimum principle, we introduce the costate variables $\bm{\pi}_x$ and $\pi_\theta$ conjugate to $\bm{x}$ and $\theta$, respectively. The system Hamiltonian reads~\citep{bryson}
\begin{equation}
    \mathcal{H} = \bm{\pi}_x \cdot \left( \bm{u} + v_{\rm s} \bm{n} \right) + \pi_\theta \left( \frac{1}{2}\omega + \Lambda S_{np}\right) \, .
    \label{eq:hamiltonian}
\end{equation}
The necessary condition for optimality is $\partial_\lambda\mathcal{H}=0$, which yields
\begin{equation}
    \frac{\partial\mathcal{H}}{\partial\lambda} = \pi_\theta S_{np}\frac{{\rm d}\Lambda}{{\rm d}\lambda}= \pi_\theta S_{np} \frac{4\lambda}{(\lambda^2+1)^2} = 0 \hspace{0.2cm}\Longrightarrow \hspace{0.2cm}\lambda=0 \hspace{0.2cm}\Longleftrightarrow\hspace{0.2cm} \Lambda=-1 \, .
\end{equation}
The optimal escape strategy, therefore, consists of maintaining a constant disk-like shape. 
In this limit, the flow-induced torque in Jeffery’s equation has the same torque structure arising in Zermelo’s minimum-time navigation problem (see discussion in Sec.~\ref{sec:baselines} and Ref.~\cite{piro2024Energetica}).

\section{Supplemental videos}
\label{app:videos}
This appendix provides a brief description of the supplemental videos accompanying the paper. The videos illustrate representative microswimmer trajectories and visualise their dispersion. 

\textbf{Supplemental Video 1.}
Trajectories of microswimmers following the naive strategy (black curve), the STO strategy (green curve), and the smart strategy (orange curve) in a two-dimensional stochastic flow at ${\rm Ku}=0.1$. The background colour represents the vorticity field. The inset shows their distance from the initial position $(x,y)=(0,0)$ as a function of time. The text in the upper-left corner indicates the instantaneous shape factor $\Lambda$ of each microswimmer. For clarity of visualization, the stochastic flow is implemented in a large domain, so no periodic boundary conditions are required.

\textbf{Supplemental Video 2.}
Same as Video 1 for ${\rm Ku}=1$.

\textbf{Supplemental Video 3.}
Same as Video 1 for ${\rm Ku}=10$.

\textbf{Supplemental Video 4.}
Same as Video 1 for ${\rm Ku}=10^7$.

\textbf{Supplemental Video 5.}
Same as Video 1 for microswimmers in turbulent flow generated by DNS. The flow field is periodically extended both in $x$ and $y$ directions. The smart strategy is trained on a stochastic flow with ${\rm Ku}=10$.

\textbf{Supplemental Video 6.}
Instantaneous positions of microswimmers following the naive strategy (black dots), the STO strategy (green dots), and the smart strategy (orange dots) in a two-dimensional stochastic flow at ${\rm Ku}=0.1$. The inset displays their ensemble-averaged distance from the initial position $(x,y)=(0,0)$ as a function of time.
For clarity of visualization, the stochastic flow is implemented in a large domain that requires no periodic boundary conditions.

\textbf{Supplemental Video 7.}
Same as Video 6 for ${\rm Ku}=1$.

\textbf{Supplemental Video 8.}
Same as Video 6 for ${\rm Ku}=10$.

\textbf{Supplemental Video 9.}
Same as Video 6 for ${\rm Ku}=10^7$.

\textbf{Supplemental Video 10.}
Same as Video 6 for microswimmers in turbulent flow generated by DNS. The flow field is periodically extended both in $x$ and $y$ directions. The smart strategy is trained on a stochastic flow with ${\rm Ku}=10$.

\end{appendix}\clearpage


%

\end{document}